\documentclass[12pt]{article}
\input psfig.sty
\usepackage{amsmath}
\usepackage{amssymb}
\usepackage{eucal}
\usepackage{graphics}

\def\nn{\noindent}

\def\Re{{\cal R \mskip-4mu \lower.1ex \hbox{\it e}\,}}
\def\Im{{\cal I \mskip-5mu \lower.1ex \hbox{\it m}\,}}
\def\ie{{\it i.e.}}

\def\etal{{\it et al.}}

\def\sub#1{_{\lower.25ex\hbox{$\scriptstyle#1$}}}

\def\to{\rightarrow}
\def\slash{\not\!}
\def\subw{_{\rm w}}
\def\mh{\ifmmode m\sbl H \else $m\sbl H$\fi}
\def\mch{\ifmmode m_{H^\pm} \else $m_{H^\pm}$\fi}
\def\mt{\ifmmode m_t\else $m_t$\fi}
\def\mc{\ifmmode m_c\else $m_c$\fi}
\def\mz{\ifmmode M_Z\else $M_Z$\fi}
\def\mw{\ifmmode M_W\else $M_W$\fi}
\def\mws{\ifmmode M_W^2 \else $M_W^2$\fi}
\def\mhs{\ifmmode m_H^2 \else $m_H^2$\fi}
\def\mzs{\ifmmode M_Z^2 \else $M_Z^2$\fi}
\def\mts{\ifmmode m_t^2 \else $m_t^2$\fi}
\def\mcs{\ifmmode m_c^2 \else $m_c^2$\fi}
\def\mchs{\ifmmode m_{H^\pm}^2 \else $m_{H^\pm}^2$\fi}
\def\ztwo{\ifmmode Z_2\else $Z_2$\fi}
\def\zone{\ifmmode Z_1\else $Z_1$\fi}
\def\mtwo{\ifmmode M_2\else $M_2$\fi}
\def\mone{\ifmmode M_1\else $M_1$\fi}
\def\tb{\ifmmode \tan\beta \else $\tan\beta$\fi}
\def\xw{\ifmmode x\subw\else $x\subw$\fi}
\def\ch{\ifmmode H^\pm \else $H^\pm$\fi}
\def\lum{\ifmmode {\cal L}\else ${\cal L}$\fi}
\def\inpb{\ifmmode {\rm pb}^{-1}\else ${\rm pb}^{-1}$\fi}
\def\infb{\ifmmode {\rm fb}^{-1}\else ${\rm fb}^{-1}$\fi}
\def\epem{\ifmmode e^+e^-\else $e^+e^-$\fi}
\def\ppb{\ifmmode \bar pp\else $\bar pp$\fi}
\def\bsg{\ifmmode B\to X_s\gamma\else $B\to X_s\gamma$\fi}
\def\bsll{\ifmmode B\to X_s\ell^+\ell^-\else $B\to X_s\ell^+\ell^-$\fi}
\def\bstt{\ifmmode B\to X_s\tau^+\tau^-\else $B\to X_s\tau^+\tau^-$\fi}

\def\sr{\ifmmode \tilde e_R\else $\tilde e_R$\fi}
\def\sl{\ifmmode \tilde e_L\else $\tilde e_L$\fi}

\newskip\zatskip \zatskip=0pt plus0pt minus0pt
\def\matth{\mathsurround=0pt}

\def\gsim{\mathrel{\mathpalette\atversim>}}
\def\atversim#1#2{\lower0.7ex\vbox{\baselineskip\zatskip\lineskip\zatskip
  \lineskiplimit 0pt\ialign{$\matth#1\hfil##\hfil$\crcr#2\crcr\sim\crcr}}}

\def\be{\begin{equation}}
\def\ee{\end{equation}}
\def\bea{\begin{eqnarray}}
\def\eea{\end{eqnarray}}
\def\pmuu{\partial^{\mu}}
\def\pmud{\partial_{\mu}}

\renewcommand{\thefootnote}{\fnsymbol{footnote}}

\def\slash{\not\!}

\begin{document}
\begin{titlepage}
\rightline{\vbox{\halign{&#\hfil\cr
&SLAC-PUB-8782\cr
&April  2002\cr}}}
\vspace{1in}
\begin{center}

{\Large\bf Supersymmetric Extra Dimensions: Gravitino Effects in Selectron
Pair Production
}\footnote{Work supported by the Department of
Energy, Contract DE-AC03-76SF00515}
\medskip

\normalsize
{\large JoAnne L. Hewett and Darius Sadri} \\
\vskip .3cm
Stanford Linear Accelerator Center \\
Stanford CA 94309, USA\\
\vskip .3cm

\end{center}


\begin{abstract}

We examine the phenomenological consequences of a supersymmetric bulk in
the scenario of large extra dimensions.  We assume supersymmetry is
realized in the bulk and study the interactions of the
resulting bulk gravitino Kaluza-Klein (KK) tower of states, with
supersymmetry breaking on the brane inducing a light mass for the
zero-mode gravitino.  We derive the 4-d effective theory, including
the couplings of the bulk gravitino
KK states to fermions and their scalar superpartners.  The
virtual exchange of the gravitino KK states in selectron pair production
in polarized \epem\ collisions is then examined.  We find that the
leading order operator for this exchange is dimension six, in
contrast to that of bulk graviton KK exchange which induces a
dimension eight operator at lowest order.  The resulting kinematic
distributions for selectron production are dramatically altered from
those in $D=4$ supersymmetric scenarios, and can lead to a  
enormous sensitivity to the fundamental higher dimensional Planck scale,
of order $20-25\times \sqrt s$.

\end{abstract}

\renewcommand{\thefootnote}{\arabic{footnote}} \end{titlepage}

\section{Introduction}

There has been much interest recently in the framework,
proposed by Arkani-Hamed, Dimopoulos, and Dvali (ADD)
\cite{Arkani-Hamed:1998rs, Antoniadis:1998ig},
which resolves the hierarchy problem by exploiting the geometry of spacetime.
In this scenario, the fundamental scale of gravity  in a higher
$D=4+\delta$ dimensional spacetime  is assumed to be of order  the
electroweak scale  $\sim 1$ TeV.   The apparent weakness of
gravity in our 4-dimensional world originates from the large  volume
of the additional $\delta$ spatial dimensions.  The
4-dimensional Planck scale is
no longer a fundamental scale, leaving the
electroweak scale as the ultraviolet cut-off
of the low-energy effective theory.  The gauge hierarchy is thus effectively
eliminated and reduced to the more tractable problem of stabilizing the
higher dimensional radii \cite{jmr}.  In this scenario, gravity propagates
throughout the higher dimensional volume, known as the bulk, whereas
the Standard Model (SM) fields are confined to a $3$-dimensional brane,
or wall.

In this theory Gauss' Law relates the Planck scale of
the 4-dimensional theory, $M_P$, to the fundamental scale of gravity, $M_D$,
through the volume of the compactified dimensions $V_\delta$ via
\begin{equation} \label{planck:relation}
        M_{P}^2  = V_\delta  M_D^{2+\delta}\,,
\end{equation}
where $M_P=1.2\times 10^{19}$ GeV is the 4-d
Planck scale.  Setting $M_D\sim 1$ TeV then determines the
compactification radius $R_c$ of the extra dimensions, with the
exact relationship being set by the geometry of the compact dimensions.
Assuming that the extra dimensions are of toroidal form  and
are all of equal size, we have $V_\delta=(2\pi R_c)^\delta$.
$R_c$ then ranges from a sub-millimeter to a few fermi for $\delta=2$ to 6.
The case of $\delta=1$ is excluded as it predicts corrections to Newtonian
gravity at distances comparable to those in
the solar system.  A similar scenario
can be realized in string theory where the string
scale  plays the role of the higher dimensional fundamental scale
\cite{Witten:1996mz},
with the string scale acting as the ultraviolet cut-off of the theory.

Proposals for the localization of SM matter and gauge fields to a $3+1$
dimensional wall have been
made in the context of topological defects of higher dimensional field
theories \cite{Rubakov:bb}.  Such localization
can occur naturally in string theory via D-branes
where the SM particles are represented by open strings whose ends lie on
the D-brane, while gravitons, which carry no gauge charges, may propagate
in the bulk and correspond to
closed strings \cite{Antoniadis:1998ig, Horava:1995qa, Polchinski:1996na}.

Since this scenario modifies gravity at the electroweak scale, it is
natural to expect the emergence of new phenomena at the TeV scale which
may reveal itself in experiments and lead to
signatures very different from SM predictions.  Upon compactification, the
bulk graviton expands into a Kaluza-Klein (KK) tower of states,
referred to as a bulk graviton KK tower,
which interact with the SM fields on the brane. Collider signals
for the graviton KK states have been studied by various authors
\cite{Giudice:1998ck,Hewett:1998sn}, who have
considered the virtual exchange of bulk graviton Kaluza-Klein towers,
processes which radiate gravitons into the bulk, and stringy
excitations of the Standard Model particles.  Data from the Tevatron,
LEPII, and HERA \cite{expt} presently constrain $M_D\gsim 1$ TeV for all
values of $\delta$,
while the LHC and  a future high energy $e^+e^-$ Linear Collider are
expected to probe fundamental scales in the $5-9$ TeV range.
Astrophysical and cosmological considerations \cite{Cullen:1999hc}
place stringent bounds, of order $M_D\gsim 100$ TeV, for the case
of $\delta=2$; these limits weaken substantially to $M_D\gsim 1$ TeV
for higher values of $\delta$.  Mechanical  experiments have tested the
inverse-square nature of the gravitational force law down to distances
of $150\mu m$ \cite{Hoyle:2000cv} for the case of $\delta=2$.  This
scenario is thus consistent
with all data as long as the number of extra dimensions is
greater than one, with the case of $\delta=2$ being disfavored in terms
of being relevant to the hierarchy problem.

While the original motivation for the ADD scenario was to solve the
hierarchy problem without the introduction of low-energy supersymmetry
(or technicolor), one still
might ask whether supersymmetry plays a role in such a
scenario.  Clearly, bulk supersymmetry is not in conflict with the
basic assumptions of the model.  In fact, various reasons exist for believing
in a supersymmetric bulk,
not least of which is the motivation of string theory.
As discussed above, D-branes of string theory provide a
natural mechanism for the confinement of the SM fields.
If string theory is the ultimate theory of nature then the proposal of ADD
might be embedded within it with a supersymmetric bulk, \ie, a
bulk supporting a supersymmetric gravitational action.
In addition, extra dimensional geometries have been shown to provide
novel methods for breaking supersymmetry \cite{susybreak,mp},
the possibility that
a supersymmetric bulk might provide a source for a tiny cosmological
constant has been discussed \cite{Schmidhuber:2000cy}, and
supersymmetry might also
serve as a mechanism for stabilizing the bulk radii.
Supersymmetry has also been considered in the context of warped
extra dimensions present in the Randall-Sundrum scenario of localized
gravity \cite{Altendorfer:2000rr}.

In this paper, we investigate the consequences of a supersymmetric
bulk in the ADD scenario.  If bulk supersymmetry
remains unbroken away from the brane,
then it is natural to ask what happens
to the superpartners of the bulk gravitons, the gravitinos.
The bulk gravitinos must also expand into a Kaluza-Klein tower of states
and induce experimental signatures.  Up to now, the
phenomenology of such gravitino KK states has been unexplored.  The
existence of a light graviphoton, which is present if there
is no orbifolding, in supersymmetric ADD
has been shown to alter the non-supersymmetric ADD phenomenology
\cite{Atwood:2000au},
and hence we also expect large effects from the gravitino sector.
Here, for concreteness, we examine the virtual exchange
of the bulk gravitino KK tower in superparticle pair production.

We now outline the approach taken in this paper.  We study the
compactification of the gravitino sector of a generic supergravity
theory living in the bulk and work out the form of the effective
action describing the free part of the reduced action on the $3+1$
dimensional brane to which the fields of the Minimal Supersymmetric
Standard Model (MSSM) are assumed to be confined. We assume that
supersymmetry on the wall is broken, giving rise to masses for the
zero-modes of the Kaluza-Klein tower resulting from the reduction of
the higher dimensional gravitinos, and shifting the masses of all
higher modes. We make no special assumptions
about the nature or origin of confinement of the MSSM fields to the
wall. We then derive the couplings of the Kaluza-Klein modes of the
higher dimensional gravitinos to fermions and their scalar
superpartners on the wall, yielding an effective 4-d theory which we then
use to study the corrections to collider signatures of certain
processes in the MSSM.
For purposes of illustration, we work with a
$10$ dimensional theory; the generality of our results
should be apparent (up to considerations of representations of
fermions in various dimensions).

The gravitino zero-mode acquires a mass from the spontaneous breaking
of supersymmetry on the brane, with this mass being proportional to
$\sim\Lambda^2_{SUSY}/M_P$, where $\Lambda_{SUSY}$ represents the
supersymmetry breaking scale in a generic model, noting that the
zero-mode gravitino couples with the usual $M_P^{-1}$ strength.
This familiar 4-d expression can also be seen to arise from 
the volume factor after integrating over the $\delta$ extra dimensions
and using Eq. (1).  Various SUSY breaking scenarios thus
yield different predictions ranging from ultralight gravitinos from
weak-scale supersymmetry breaking models, light gravitinos 
in gauge mediated SUSY breaking, and heavier gravitinos
in models of gravity and anomaly mediated SUSY
breaking.  The possibility of light gravitinos is an interesting one
and their generic collider \cite{coll_grav} and astrophysical
\cite{astro_grav} implications have been studied, with direct collider
searches yielding a bound of $m_{3/2}\gsim 10^{-5}$ eV \cite{expt_grav}.
In particular, the case of gauge mediated SUSY breaking (GMSB)
has been intensively studied \cite{Giudice:1998bp}, as a light
gravitino
modifies the standard collider search techniques for supersymmetry
and  provides a good candidate for warm dark
matter \cite{hitoshi}.  In this paper,
we work in the context of gauge mediated
supersymmetry as it naturally affords a light gravitino state, however
our results apply to any supersymmetric model with a light gravitino, 
including weak-scale breaking models.

We focus on the effects of the virtual exchange of
the bulk gravitino and graviton KK tower states in the process
$e^+e^-\to\tilde e^+\tilde e^-$
at a high energy Linear Collider (LC).  This process is well-known
as a benchmark for
collider supersymmetry studies \cite{Baer:1988kx},
as the use of incoming polarized beams
enables one to disentangle the neutralino sector and determine the
degree of mixing between the various pure gaugino states.
The effects of the virtual exchange of a bulk graviton KK tower
in selectron pair production has been examined in \cite{tgr} for
the case of non-supersymmetric large extra dimensions.  Here,
we will
see that the introduction of bulk gravitino KK exchange greatly alters the
phenomenology of this process by modifying the angular distributions
and by substantially
increasing the magnitude of the cross section.  We find that the
leading order behavior for this
process is given by a dimension-6 operator, in contrast to the
dimension-8 operator corresponding to graviton KK exchange.  This
yields
a tremendous sensitivity to the existence of a supersymmetric bulk,
resulting in a search reach for the ultraviolet cut-off of the theory
of order $20-25\times\sqrt s$.

Our paper is organized as follows:
In Section 2 we present the Kaluza-Klein decomposition of the
gravitinos in such a model. In Section 3 we use the general
Noether technique to couple the gravitino Kaluza-Klein states to
matter fields, and in Section 4 we present the
phenomenology of this scenario, pointing out some generic features
of such models.  Our conclusions are given in Section 5.
Various details are relegated to the appendices.

Our notation in this paper is as follows: We assume space-time
possesses $3+1+\delta$ dimensions, with matter confined to a $3+1$
dimensional brane, and pure supergravity propagating also in the
extra $\delta$ dimensions.
Letters from the Greek alphabet are used to denote curved (world)
indices while those from the Latin alphabet denote flat (tangent space)
indices.  Hatted symbols range over the full $4 + \delta$ dimensions;
barred ones are restricted to the $\delta$ extra dimensions,
while those without any decoration live on the $3+1$ dimensional brane.
Coordinates over the entire $D=4+\delta$ dimensional manifold split as
$z^{\hat{\mu}} \: = \: \left( x^{\mu} , y^{\bar{\mu}} \right)$.
Our Minkowski metric convention is
$\eta_{\hat{a} \hat{b}} \: = \: ( +1,-1,\ldots,-1 )$.

\section{Kaluza-Klein Excitations of the Gravitino}
\label{Kaluza-Klein Excitations of the Gravitino}

We begin by reminding the reader of some general observations about
string theory and supergravity in ten and eleven dimensions.
Recall that type IIA (ten dimensional) supergravity with two supersymmetries of
opposite chirality can be deduced via dimensional reduction
of the unique 11 dimensional (N=1) supergravity theory.
The type IIA superstring theory whose low energy effective
action gives rise to the type IIA supergravity
admits Dp-branes with p even, and is
S-dual to eleven dimensional
M-theory \cite{Polchinski:rq}, with the eleven dimensional
supergravity as its low energy limit.
M-theory is believed to unify the five known string
theories.
The type IIB string theory (which admits Dp-branes with p odd) is
T-dual to the type IIA theory.

D-branes are extended string theoretic objects on which open
strings terminate \cite{Polchinski:1996na}.
Only closed strings can propagate far away from
the D-brane, which on a ten dimensional background is
described locally by a type II string theory whose spectrum
contains two gravitinos (their vertex operators carry one vector and
one spinor index).
The D-brane introduces open string boundary conditions which are
invariant under just one supersymmetry, so only a linear
combination of the two original supersymmetries survives for the open
strings. Since open and closed strings
couple to each other, the D-brane breaks the original $N=2$
supersymmetry down to $N=1$. The low energy effective theory will then be a
$D=10, N=1$ supergravity theory with a single Majorana-Weyl gravitino
which couples to a
conserved space-time supercurrent. For our model, we assume that the
Standard Model fields are confined to a 3-brane, with pure
supergravity living in the bulk of space-time.
Attempts at constructing a Standard Model on D-branes can be found in
\cite{Berenstein:2001nk}.

Pure supergravity in four dimensions contains as physical fields the
vierbein $e^{\: \: \: m}_{\mu}$ and the gravitino $\Psi_{\mu}$
\cite{VanNieuwenhuizen:ae, Freedman:1976xh}.
In the free
field limit, the gauge action of supergravity reduces to the standard
Fierz-Pauli action (the linearized part of the Einstein-Hilbert
action), together with the Rarita-Schwinger action, which are
the unique ghost free actions for a spin $2$ and spin $\frac{3}{2}$
field \cite{VanNieuwenhuizen:fi, Rarita:mf}.

The spin of a field $\Phi(z)$ is given by a representation of the tangent
space group at a fixed point $z$ \cite{Green:sp, Salam:1981xd}.
Under the above compactification,
the metric $g_{\hat{\mu} \hat{\nu}}$ decomposes in four dimensions as
a metric $g_{\mu \nu}$, while the components $g_{\mu \bar{\nu}}$ with
one index ranging over the extra dimensions are seen in four
dimensions as a vector, and $g_{\bar{\mu} \bar{\nu}}$ transforms as a
scalar.
To study the behavior of spinors on a general manifold, we must
introduce a set of orthonormal basis vectors (the vielbein)
$e_{\hat{\mu}}^{\: \: \: \hat{m}}(z)$
in the tangent space of
the manifold at the point $z$, with $\hat{\mu}$ transforming
as a vector index under general coordinate changes, while $\hat{m}$
transforms as a scalar under such diffeomorphisms, but as a local
Lorentz vector index under $SO(D-1,1)$, i.e., the Lorentz group acting
in the tangent space at the point $z$, satisfying
\cite{Birrell:ix, Weinberg:cr}
\begin{equation}
\label{defn:vierbein}
   \eta^{{\hat{m} \hat{n}}} \: = \: g^{\hat{\mu} \hat{\nu}}
          e^{\: \: \: \hat{m}}_{\hat{\mu}} e^{\: \: \: \hat{n}}_{\hat{\nu}}\,,
   \: \: \: \: \: \: \: \: \: \: \: \: \: \: \: \: \: \: \: \:
   g_{{\hat{\mu} \hat{\nu}}} \: = \: \eta_{\hat{m} \hat{n}}
          e^{\: \: \: \hat{m}}_{\hat{\mu}} e^{\: \: \: \hat{n}}_{\hat{\nu}}\,.
\end{equation}
Associated with the local Lorentz symmetry is a gauge field for local
Lorentz transformations, the spin connection $\omega_{\hat{\mu} \hat{m}
\hat{n}}$, transforming as a covariant vector under general coordinate
transformations.

We now assume the vacuum of space-time to be of the form
$M^{4} \times T^{6}$, where $M^{4}$ is four dimensional
Minkowski space-time and $T^{6} = S^{1} \times \ldots \times S^{1}$,
the direct product of six dimensions each
compactified on a circle. This product form of the vacuum insures four
dimensional Poincar\'{e} invariance. Physical fluctuations of
space-time can be treated in perturbation theory by expanding the
metric around this vacuum.

Consistent with the symmetry of the vacuum state,
we assume the vielbein takes the form
\begin{equation} \label{vierbein}
  E_{\hat{m}}^{\: \: \: \hat{\mu}} \: = \:
  \left(
  \begin{matrix}
    E_{m}^{\: \: \: \mu} (x)
           & \sum_{a} A_{\bar{m}}^{a} (x) \: K_{\bar{m}}^{a} (y)\\
    0 & e_{\bar{m}}^{\: \: \: \bar{\mu}} (y)
  \end{matrix}
  \right) \,,
\end{equation}
where we have used the local SO(9,1) gauge symmetry to set the
$E_{\bar{m}}^{\: \: \: \mu}$ components to zero. Here
$e_{\bar{m}}^{\: \: \: \bar{\mu}}$ is the vielbein on the compactified
space, $A_{\bar{m}}^{a}$ are the massless gauge fields of the symmetry
group of the internal space (here $U(1)^{6}$), and the
$K_{\bar{m}}^{a}$ are the Killing vectors associated with this symmetry.

The gravitino kinetic term of the Lagrangian is
\begin{equation} \label{RS:action}
    E^{-1} \mathcal{L} \: = \:
         \frac{i}{2} \: \bar\Psi_{\hat{\mu}} \:
         \Gamma^{\hat{\mu}\hat{\nu}\hat{\rho}}
         \: \triangledown_{\hat{\nu}}
         \Psi_{\hat{\rho}} \,,
\end{equation}
with $E$ being the determinant of the vielbein in ten dimensions,
$\Gamma^{\hat{\mu}\hat{\nu}\hat{\rho}}$ the antisymmetric product of three
$\Gamma$ matrices defined by
\begin{equation}
  \Gamma^{\hat{\mu}\hat{\nu}\hat{\rho}} \: = \:
  \Gamma^{[ \hat{\mu}\hat{\nu}\hat{\rho} ]} \: = \:
  \frac{1}{3 !}
  \left(
  \Gamma^{\hat{\mu}} \Gamma^{\hat{\nu}} \Gamma^{\hat{\rho}} \: - \:
  \Gamma^{\hat{\nu}} \Gamma^{\hat{\mu}} \Gamma^{\hat{\rho}} \: + \:
  \ldots \right)\,,
\end{equation}
and $\Psi_{\hat{\mu}}$ a Majorana-Weyl vector-spinor.
We have suppressed the spinor indices for notational clarity.

We use $\Gamma^{\hat{\mu}}$ to denote Dirac gamma matrices for the full $D$
dimensional space, and $\gamma^{\mu}$ to denote them on the $3+1$
dimensional brane.
As usual, the Dirac algebra (in the tangent space)
is spanned by a set of constant
matrices (which in $10$ dimensions are $32 \times 32$)
\begin{equation} \label{clifford:algebra:flat}
  \{ \Gamma^{\hat{m}} , \Gamma^{\hat{n}} \}
  \: = \: 2 \: \eta^{\hat{m} \hat{n}}\,.
\end{equation}
The curved space Dirac matrices are field dependent and given by
\begin{equation}
  \Gamma^{\hat{\mu}}(z) \: = \:
  e^{\hat{\mu}}_{\: \: \: \hat{m}}(z) \Gamma^{\hat{m}}
\end{equation}
satisfying
\begin{equation} \label{clifford:algebra:curved}
  \{ \Gamma^{\hat{\mu}}(z) , \Gamma^{\hat{\nu}}(z) \}
  \: = \: 2 \: g^{\hat{\mu} \hat{\nu}}(z)\,.
\end{equation}

A convenient representation of the algebra \eqref{clifford:algebra:flat} in
ten dimensions which makes manifest
the four dimensional decomposition is given in Appendix \ref{appendix:rep}.
The SO(9,1) generators are
\begin{equation} \label{so91:gen}
  \Sigma^{\hat{m} \hat{n}} \: = \:
  \frac{i}{4} \: \left[ \Gamma^{\hat{m}} , \Gamma^{\hat{n}} \right]\,.
\end{equation}
The first four matrices in the set \eqref{10d:gamma:matrices}
furnish a reducible representation
of $SO(3,1)$. Taking account of the identity acting in the internal
$2^{\frac{\delta}{2}} = 8$ dimensional spinor space,
we see that a $D=10$ Dirac spinor
decomposes into $8$ $D=4$ Dirac spinors. A $D=10$ Majorana spinor
would decompose into $8$ Majorana spinors in four dimensions, and a
$D=10$ Majorana-Weyl spinor would decompose into four Majorana
spinors, since the $D=10$ chirality condition $\Gamma_{11} \Psi = \pm
\Psi$ (where $\Gamma_{11} = i \Gamma_0 \ldots \Gamma_9$ plays the role
in ten dimensions of $\gamma_5$ in four dimensions)
would pair-wise relate half the degrees of freedom, leaving four $D=4$
Majorana spinors.
There will also be
24 Majorana
spin-$\frac{1}{2}$ fields.
After this reduction we recover the standard definition of
the four dimensional spinor generator.
The metric in ten
dimensions decomposes into one spin-$2$ graviton in four dimensions, 6
vectors, and 21 scalars.

In ten dimensions, a Majorana-Weyl spinor contains $16$ real
components.  To see this, note that in ten dimensions, the Dirac
matrices are $32 \times 32$, so a Dirac spinor would contain $32$
complex components.  The Majorana condition reduces this in half by
imposing a reality condition.  Finally, the Weyl condition reduces this
again, for a final total of 16 real components. The equations of motion
for this spinor imply that not all of the remaining components
are independent, reducing further the independent propagating degrees
of freedom.

For simplicity we assume the
compactification radii of all six dimensions to be the same, though
the generalization is obvious. Physically this compactification
amounts to the identification
\begin{equation}
  y^{\bar{\mu}} \: = \: y^{\bar{\mu}} + 2 \pi n R_{c}\,,
\end{equation}
with $n$ an arbitrary integer and
$R_{c}$ the common radius of the compactified dimensions.
The condition on the fields then becomes, recalling our notation
$z^{\hat{\mu}} \: = \: \left( x^{\mu} , y^{\bar{\mu}} \right)$,
\begin{equation}
  \Psi_{\hat{\mu}}
    ( x^{\nu}, y^{\bar\nu} + 2 \pi n R_{c} )
         \: = \:
  \Psi_{\hat{\mu}}
    ( x^{\nu}, y^{\bar\nu} )\,.
\end{equation}

The transformation properties of the Rarita-Schwinger
field $\Psi_{\hat{\mu}}$ are the product of a vector and a spinor.
The covariant derivative $\triangledown_{\hat{\mu}}$ is given by
\begin{equation}
  \triangledown_{\hat{\mu}} \: = \: \partial_{\hat{\mu}} \: + \:
  \frac{i}{4} \omega_{\hat{\mu} \hat{m} \hat{n}}
  \Sigma^{\hat{m} \hat{n}}\,.
\end{equation}
where $\Sigma^{\hat{m} \hat{n}}$ are the SO(9,1) generators given in
\eqref{so91:gen},
and $\omega_{\hat{\mu} \hat{m} \hat{n}}$ is a sum of the standard
spin connection and terms quadratic in gravitino fields
\cite{Cremmer:1978km}, with
supercovariantization giving rise to four fermion terms.

Using the vielbein to translate from curved to flat indices, we can
rewrite the action \eqref{RS:action}, after linearizing the spin
connection, in the form
\begin{equation} \label{RS:action:flat}
    E^{-1} \mathcal{L} \: = \:
         \frac{i}{2} \: \bar\Psi_{\hat{m}} \:
         \Gamma^{\hat{m}\hat{n}\hat{r}}
         \: \partial_{\hat{n}}
         \Psi_{\hat{r}}\,,
\end{equation}
where
\begin{subequations}
\begin{align}
  \Psi_{\hat{m}} \: &= \:
  E^{\: \: \: \hat{\mu}}_{\hat{m}}
  \Psi_{\hat{\hat{\mu}}}\,,
\end{align}
\begin{align}
  \partial_{\hat{n}} \: &= \:
  E^{\: \: \: \hat{\nu}}_{\hat{n}}
  \partial_{\hat{\nu}} \,.
\end{align}
\end{subequations}
The linearized action possesses the local Abelian gauge symmetry
\begin{equation} \label{gauge:invariance}
  \delta \Psi_{\hat{m}}
  \: = \: \partial_{\hat{m}} \epsilon\,,
\end{equation}
with $\epsilon$ a Majorana-Weyl spinor. This is the same linearized
gauge invariance one would expect of the
fermionic part of the gauge action of supergravity.
The above invariance follows from the
realness of $\mathcal{L}$ and the total antisymmetry of
$\Gamma^{\hat{m} \hat{n} \hat{p}}$.

To make the Kaluza-Klein decomposition
\cite{Salam:1981xd}, we expand the Rarita-Schwinger
field in eigenfunctions of the compactified space (of dimension
$\delta$, which for now is taken to be arbitrary),
assumed here to be the torus $T^{\delta}$, with volume
$V_{\delta} = (2 \pi R_{c})^{\delta}$.
The expansion is
\begin{equation} \label{expansion}
  \Psi_{\hat{m}} (z) \: = \:
  \sum_{\vec{n}=-\infty}^{\infty}
  \frac{\Phi_{\hat{m}}^{\vec{n}}(x)}{\sqrt{V_{\delta}}} \:
  e^{\frac{i \vec{n} \cdot \vec{y}}{R_{c}}}\,.
\end{equation}
The four dimensional fields are seen to arise as the coefficients in this
expansion.

We take as our starting point the linearized action
\eqref{RS:action:flat}.
We now substitute \eqref{expansion}
together with \eqref{vierbein}
in the action \eqref{RS:action:flat}, then decompose the 10 dimensional
indices as $\hat{m} = (m , \bar{m})$, with the indices $m$ and $\bar{m}$
transforming as vectors under $SO(3,1)$ and $SO(\delta)$ respectively.
The Rarita-Schwinger field then splits as
\begin{equation}
  \Psi_{\hat{m}} \: = \: ( \Psi_{m} , \Psi_{\bar{m}} )\,,
\end{equation}
where
\begin{subequations}
\begin{align}
  \Psi_{m} \: &= \:
  E^{\: \: \: \mu}_{m}
  \Psi_{\mu}\,,
\end{align}
\begin{align}
  \Psi_{\bar{m}} \: &= \:
  E^{\: \: \: \bar{\mu}}_{\bar{m}}
  \Psi_{\bar{\mu}} \: = \:
  e^{\: \: \: \bar{\mu}}_{\bar{m}}
  \Psi_{\bar{\mu}}\,. 
\end{align}
\end{subequations}
Introducing the shorthand notation
\begin{equation}
  \alpha \: = \: \frac{- i \vec{s} \cdot \vec{y}}{R_{c}}\,,
  \: \: \: \: \: \: \: \: \: \: \: \: \: \: \: \: \: \: \: \:
  \beta  \: = \: \frac{+ i \vec{t} \cdot \vec{y}}{R_{c}} \: \: \,,
\end{equation}
we find
\begin{equation} \label{decomposition:1}
\begin{split}
e^{-1} \mathcal{L}
\:  = \:
  \frac{i}{2 V^{\delta}} \:
  \sum_{\vec{s}=-\infty}^{\infty} \:
  \sum_{\vec{t}=-\infty}^{\infty} \:
  \Big\{ 
&  \Big(
      \bar{\Phi}_{m}^{\vec{s}}
      \Gamma^{mnp} ( \partial_{n} \Phi_{p}^{\vec{t}} )
                \: + \:
      \bar{\Phi}_{m}^{\vec{s}}
      \Gamma^{mn\bar{p}} ( \partial_{n} \Phi_{\bar{p}}^{\vec{t}} ) \\
                \: + \:
&     \bar{\Phi}_{\bar{m}}^{\vec{s}}
      \Gamma^{\bar{m}np} ( \partial_{n} \Phi_{p}^{\vec{t}} )
                \: + \:
      \bar{\Phi}_{\bar{m}}^{\vec{s}}
      \Gamma^{\bar{m}n\bar{p}} ( \partial_{n} \Phi_{\bar{p}}^{\vec{t}} )
  \Big)
  \: e^{\alpha + \beta} \\
                \: + \:
& \Big(
      \bar{\Phi}_{m}^{\vec{s}}
      \Gamma^{m \bar{n}p} ( \partial_{\bar{n}} e^{\beta} )
                \Phi_{p}^{\vec{t}}
                \: + \:
      \bar{\Phi}_{m}^{\vec{s}}
      \Gamma^{m \bar{n}\bar{p}} ( \partial_{\bar{n}} e^{\beta} )
      \Phi_{\bar{p}}^{\vec{t}} \\
                \: + \:
&     \bar{\Phi}_{\bar{m}}^{\vec{s}}
      \Gamma^{\bar{m} \bar{n} p} ( \partial_{\bar{n}} e^{\beta} )
      \Phi_{p}^{\vec{t}}
                \: + \:
      \bar{\Phi}_{\bar{m}}^{\vec{s}}
      \Gamma^{\bar{m} \bar{n} \bar{p}} ( \partial_{\bar{n}} e^{\beta} )
      \Phi_{\bar{p}}^{\vec{t}}
    \big) \: e^{\alpha}
  \Big\}
\end{split}
\end{equation}
with $e$ being the determinant of the four
dimensional vierbein.
The first set of terms proportional to $e^{\alpha + \beta}$ and with
the derivative acting on the fields $\Phi_{p}^{\vec{t}}$ and
$\Phi_{\bar{p}}^{\vec{t}}$
are the kinetic terms since the derivative is taken with respect to
coordinates in $M^{4}$,
while the remaining terms proportional to $e^{\alpha}$ with the
derivative acting on the exponential $e^{\beta}$
will give rise to the mass terms
of the spin $\frac{3}{2}$ and $\frac{1}{2}$ fields in the effective
four dimensional action. It is important to note that fields
with indices ranging over the $3+1$ dimensions ($\Phi_{m}$) appear in this
Lagrangian coupled to those with indices in the extra $\delta$
dimensions ($\Phi_{\bar{m}}$),
in both the kinetic and mass terms. This mixing prevents us from interpreting
$\Phi_{m}$ as the four dimensional Rarita-Schwinger field.
We introduce the field redefinition \cite{Cremmer:1979up,Fogleman:hs}
\begin{equation} \label{field:redef}
  \chi_{m}^{\vec{n}} (x) \: \equiv \: \Phi_{m}^{\vec{n}} (x) \: + \: \frac{1}{2}
  \Gamma_{m} \Gamma^{\hat{r}} \Phi_{\hat{r}}^{\vec{n}} (x)\,,
\end{equation}
which will be identified as the four dimensional Rarita-Schwinger
field.
In what follows, we shall systematically drop all terms coupling only
fields with vector
indices in the extra dimensions (these would give rise to
spin $\frac{1}{2}$ terms in the four dimensional effective theory),
and concentrate on the gravitino (spin $\frac{3}{2}$) part of the
effective theory.
The field redefinition \eqref{field:redef} completely decouples the
kinetic term into spin $\frac{1}{2}$ and spin $\frac{3}{2}$ pieces,
but the mass term still contains couplings between
spin $\frac{3}{2}$ and $\frac{1}{2}$ fields.
The mixing with spin $\frac{1}{2}$ in the mass term is due to the fact
that the field $\chi_{m}^{\vec{n}}$ transforms as the
\begin{equation}
  \left( \frac{1}{2} , \frac{1}{2} \right) \otimes
  \Bigg[ \left( \frac{1}{2} , 0 \right) \oplus
  \left( 0 , \frac{1}{2} \right) \Bigg] \: = \:
  \left( \frac{1}{2} , 1 \right) \oplus
  \left( 1 , \frac{1}{2} \right) \oplus
  \left( \frac{1}{2} , 0 \right) \oplus
  \left( 0 , \frac{1}{2} \right)
\end{equation}
representation of SO(1,3). To project out the
$(\frac{1}{2},0) \oplus (0,\frac{1}{2})$ part we impose the SO(3,1)
invariant gauge condition
\begin{equation}
  \Gamma^{m} \chi_{m}^{\vec{n}} (x) \: = \: 0\,,
\end{equation}
leaving only a spin $\frac{3}{2}$ field.

To evaluate the derivatives in the mass terms of
\eqref{decomposition:1},
we note that they are taken over coordinates in the extra
dimensions, and since $\vec{y}$ ranges over the same coordinates, they
yield
\begin{equation} \label{partial}
  \partial_{\bar{n}} e^{\beta} \: = \:
  \frac{i}{R_{c}} t_{\bar{n}} e^{\beta}\,.
\end{equation}
Putting together \eqref{decomposition:1},
\eqref{field:redef} and \eqref{partial}, we see that the spin
$\frac{3}{2}$ part of the Lagrangian takes the form
\begin{equation} \label{decomposition:2}
  e^{-1} \mathcal{L}(x,y)
   = 
  \frac{i}{2 V^{\delta}} 
  \sum_{\vec{s}} 
  \sum_{\vec{t}} 
  e^{\alpha + \beta}
  \Big\{
      \bar{\chi}_{m}^{\vec{s}}(x)
      \Gamma^{mnp} ( \partial_{n} \chi_{p}^{\vec{t}}(x) ) - 
      \bar{\chi}_{m}^{\vec{s}}(x)
                (2 \Sigma^{m p} ) \Gamma^{\bar{n}} 
                ( \frac{t_{\bar{n}}}{R_{c}}  )
                \chi_{p}^{\vec{t}}(x)
  \Big\}
\end{equation}
where
$\bar{\chi}_{m}^{\vec{s}} =
(\chi_{m}^{\vec{s}})^{\dagger} \gamma^{0} \otimes 1_{8}$,
the $\Sigma^{m n}$ are given by restricting the indices of the
SO(9,1) generators \eqref{so91:gen} to the first $3+1$ dimensions,
and we have summed over $\bar{n}$ ranging over the extra dimensions.

The only dependence on the coordinates of the extra dimensions
in \eqref{decomposition:2} is isolated in the exponential
$e^{\alpha + \beta}$.
Integrating over the extra dimensions gives the four dimensional
effective Lagrangian
\begin{equation} \label{define:eff:Lagrangian}
  \mathcal{L}_{eff} ( x ) \: = \: \int dy_{1} \ldots dy_{\delta}
  \: \mathcal{L} ( x , y )\,.
\end{equation}
The integrals in \eqref{define:eff:Lagrangian}
can be performed using the orthogonality of the exponentials, with the
result
\begin{equation}
  \int_{- \pi R_{c}}^{\pi R_{c}} dy_{1} \ldots
  \int_{- \pi R_{c}}^{\pi R_{c}} dy_{\delta} \:
  \exp
    \left[
           - \frac{i}{R_{c}} ( \vec{s} - \vec{t} ) \cdot \vec{y}
    \right]
  \: = \: V_{\delta} \prod_{j=1}^{\delta} \: \delta_{s_{j} t_{j}}\,,
\end{equation}
where $V_{\delta} = (2 \pi R_{c})^{\delta} $
is the volume of the compactified extra dimensions.
The effective Lagrangian defined by \eqref{define:eff:Lagrangian}
can be written as a sum over Kaluza-Klein states
\begin{equation}
  \mathcal{L}_{eff} ( x )
  \: = \:
  \sum_{\vec{s}=-\infty}^{\infty}
  \mathcal{L}_{eff}^{\vec{s}} ( x )\,,
\end{equation}
with the Lagrangian for each level
\begin{equation}
\label{effective:Lagrangian:KK}
  e^{-1} \: \mathcal{L}_{eff}^{\vec{s}} ( x ) \: = \:
  \frac{i}{2} \:
      \bar{\chi}_{m}^{\vec{s}}(x)
      \Gamma^{mnp} ( \partial_{n} \chi_{p}^{\vec{s}} (x))
                \: - \: i
      \bar{\chi}_{m}^{\vec{s}}(x)
                \Sigma^{m p} ( \frac{1}{R_c} \Gamma^{\bar{n}} s_{\bar{n}} )
                \chi_{p}^{\vec{s}}(x)\,,
\end{equation}
and the fields $\chi_{m}^{\vec{s}}$ describing an individual
Kaluza-Klein mode.
In \eqref{effective:Lagrangian:KK},
the $\Gamma^{mnp}$ and $\Sigma^{mp}$ are still $32 \times 32$.
We note that masses of the Kaluza-Klein excitations of the gravitino are
the same as for the gravitons.
To arrive at a proper gravitino mass term, we must diagonalize the term
$\Gamma^{\bar{n}} s_{\bar{n}}$ in the internal space.
First we note that in the representation \eqref{10d:gamma:matrices}
\begin{equation}
\begin{aligned}
  \Gamma^{mnp} \: & = \: \gamma^{mnp} \otimes 1_{8}\,, \\
  \Sigma^{mn} \: & = \: \sigma^{mn} \otimes 1_{8}\,,
\end{aligned}
\end{equation}
with the $\gamma^{mnp}$ and $\sigma^{mn}$ being products of $4 \times 4$
four dimensional Dirac matrices.
We decompose the fermions into four dimensional ones
$\chi_{\tilde{\alpha}} \rightarrow \omega_{\alpha,j}$ with
$\tilde{\alpha} = 1 \ldots 32$ and $\alpha = 1 \ldots 4$
spinor indices in ten and four dimensions respectively, and
$j = 1 \ldots 8$ serving as an internal index, and then use
$\bar{\chi}_{m}^{\vec{s}} =
(\chi_{m}^{\vec{s}})^{\dagger} \gamma^{0} \otimes 1_{8}$.
Making a unitary transformation on the fields allows us to
diagonalize $\Gamma^{\bar{n}} s_{\bar{n}}$ without affecting the
kinetic term since in the chosen representation, it is already
diagonal in the internal space.
After diagonalization in the internal subspace, the Lagrangian can be
written as
\begin{equation} \label{effective:Lagrangian:KK:diag}
  e^{-1} \: \mathcal{L}_{eff}^{\vec{s}} ( x ) \: = \:
  \sum_{j=1}^{4}
  \left\{
  \frac{i}{2} \:
      \bar{\omega}_{m}^{\vec{s},j}(x)
      \gamma^{mnp} ( \partial_{n} \omega_{p}^{\vec{s},j} (x) )
                \: + \:
      \bar{\omega}_{m}^{\vec{s},j}(x)
                \sigma^{mp} \gamma^{5} m_{\vec{s}}^j
                \omega_{p}^{\vec{s},j}(x)
  \right\}\,,
\end{equation}
with the mass eigenvalue coming from the diagonalization and is given by
$m_{\vec{s}}^j = (-1)^j \frac{\sqrt{\vec{s} \cdot \vec{s}}}{R_{c}}$.
The fields associated with the negative mass eigenvalues can be redefined
to remove this sign, however, care must be taken with the Feynman rule for
the coupling of the gravitino to matter.
The $\gamma^5$ in the mass term can be removed via a chiral rotation of the
fields, which introduces an extra factor of $i$ into the mass without
affecting the kinetic term.
The sum in \eqref{effective:Lagrangian:KK:diag}
 runs over the four Majorana vector-spinors
in four
dimensions. We have applied the Majorana-Weyl condition in ten dimensions,
which, as previously discussed, yields four Majorana spinors after
the decomposition into four dimensions.
Generally, the masses of the four gravitinos at each Kaluza-Klein
level can be shifted by supersymmetry breaking effects on the brane.
When we consider the phenomenology of such models,
we will assume that the $N=4$ supersymmetry
is broken at scales near the fundamental scale $M_D$, with only $N=1$
supersymmetry surviving down to the electroweak scale.
The phenomenological contributions from the heavy gravitinos associated
with the breaking of the extended supersymmetry
near the fundamental scale will be highly suppressed,
due to the large mass for these individual excitations.








Using the identity
\begin{equation}
  \gamma^{\mu \nu \rho} \: = \: - i \epsilon^{\mu \nu \rho \sigma}
  \gamma_{\sigma} \gamma_{5}\,,
\end{equation}
the compactified action for each gravitino
can be put in the standard form for the
action of a massive spin $\frac{3}{2}$ particle in four dimensions
\begin{equation} \label{action:standard:form}
  e^{-1} \:
  \mathcal{L}_{eff}^{\vec{s,j}} ( x ) \: = \:
  - \frac{1}{2}
  \epsilon^{m n r p}
    \bar{\omega}_{m}^{\vec{s},j}(x)
    \gamma_{5} \gamma_{n} \partial_{r} \: \omega_{p}^{\vec{s},j}(x)
  \: - \:
  \frac{m_{\vec{s}}^j}{4}  \:
  \bar\omega_{m}^{\vec{s},j}(x)
  \left[ \gamma^{m} , \gamma^{p} \right] \omega_{p}^{\vec{s},j}(x)
  .
\end{equation}
The mass term appearing in \eqref{action:standard:form}
has the same form as the mass
generated by spontaneous breaking of supersymmetry.
The massless limit of the propagator associated with this
Lagrangian contains singular terms,
but these terms are
all proportional to $q^m$ or $q^p$, and since $\omega_m$
couples to a conserved current (the Noether current associated with
supersymmetry, hence the interpretation as the gauge
field of supersymmetry), their contributions vanish.
We assume that supersymmetry on the brane is broken, giving
a mass to the lowest lying Kaluza-Klein states ($\vec{s}=0$)
in \eqref{action:standard:form},
and additively shifting the masses of the higher
modes by $m_{\vec n} \to m_{\vec n} + m_0$.
A natural mechanism for breaking supersymmetry which ensures a light
gravitino is gauge-mediation \cite{Giudice:1998bp}. In these models,
the gravitino is the lightest supersymmetric particle, and if R parity
is a good symmetry, it is the terminus for the the decay chains of all
superparticles. The lightness of the gravitino simplifies the
summation over the Kaluza-Klein states (see Appendix C)
leading to results which are essentially independent of the mass of the
zero-mode.

The Lagrangian \eqref{action:standard:form} implies the field equation
\begin{equation} \label{field:equation}
  \epsilon^{m n r p}
    \gamma_{5} \gamma_{n} \partial_{r} \: \omega_{p}^{\vec{s},j}(x)
  \: + \:
  \frac{m_{\vec{s}}^j}{2}  \:
  \left[ \gamma^{m} , \gamma^{p} \right] \omega_{p}^{\vec{s},j}(x)
  \: = \: 0
\end{equation}
obeyed by each Kaluza-Klein excitation.
Contraction of \eqref{field:equation} with $\gamma_{m}$ yields
\begin{equation}
  \gamma^{m} \: \omega_{m}^{\vec{s},j} \: = \: 0\,,
\end{equation}
since by taking the divergence of \eqref{field:equation}, we have
$\left[ \slash{\partial} , \gamma^{n} \right] \omega_{n}^{\vec{s},j}=0$.
This shows that the field equation \eqref{field:equation} indeed
describes a particle of spin $\frac{3}{2}$ with no spin $\frac{1}{2}$
admixture \cite{Weinberg:cr}.
Each vector component ($m$) of the
Rarita-Schwinger field $\omega_{m}^{\vec{s},j}$
satisfies a Dirac equation.

We note that the kinetic part of the compactified action
still possesses the local gauge symmetry
arising from the decomposition of \eqref{gauge:invariance},
but the mass term breaks this
symmetry, thus allowing us to invert the free quadratic operator in
the Lagrangian to find an appropriate propagator, which we derive in
Appendix B.





\section{Gravitino Couplings to Matter}



In this section we discuss the coupling of fermions and scalars to gravitinos.
We begin by formulating a globally supersymmetric theory,
and then proceed to gauge this supersymmetry via the Noether
procedure, deriving a locally supersymmetric Lagrangian yielding the
coupling of the fermions and scalars to the graviton and gravitino.
The Noether procedure provides a systematic technique for deriving an
action with a local symmetry from an action possessing a global one
\cite{Freedman:1976xh,Ferrara:1983gn}.
Gauging  the rigid supersymmetry transformations yields the
coupling of the Rarita-Schwinger field, which is the
gauge field of the local supersymmetry, to the matter multiplet.


The $\gamma$ matrices are, in the chiral representation, given
by
\begin{equation}
    \gamma^{\mu} \: = \:
    \left(
    \begin{matrix}
        0 & \sigma^{\mu} \\
        \bar{\sigma}^{\mu} & 0
    \end{matrix}
    \right)\,,
\end{equation}
with
\begin{equation}
  \sigma^{\mu} \: = \:
    \left(
           1 , \vec{\sigma}
    \right)\,,
  \: \: \: \: \: \: \: \: \: \: \: \: \: \: \:
  \bar{\sigma}^{\mu} \: = \:
    \left(
           1 , - \vec{\sigma}
    \right)\,,
\end{equation}
and $\vec{\sigma}$ being the Pauli matrices. A Dirac spinor
is constructed as follows:
\begin{equation} \label{dirac:spinor}
    \Psi \:= \:
    \left(
    \begin{array}{c}
        \psi_{L} \\
        \psi_{R}
    \end{array}
    \right) \: = \:
    \left(
    \begin{array}{c}
        \psi_{\alpha} \\
        \bar{\chi}^{\dot{\beta}}
    \end{array}
    \right)\,,
\end{equation}
where we have introduced dotted and undotted indices \cite{Bagger:1990qh}, and
$\psi_{L/R}$ are, in the massless limit, states of definite helicity each satisfying
a Weyl equation.

We begin with the following off-shell Lagrangian in $3+1$ dimensions
(two copies of $N=1$, $D=4$, chiral supermultiplet)
\begin{equation}
    \label{matter:Lagrangian}
    \mathcal{L} \: = \:
    ( \partial_{\mu} \Phi^{\dagger} )
    ( \partial^{\mu} \Phi ) \: + \:
    i \: \bar{\Psi} \gamma^{\mu} \partial_{\mu} \Psi
    \: + \: \mathcal{F}^{\dagger} \mathcal{F}\,,
\end{equation}
which describes a
free massless fermion, a multiplet of complex scalar fields,
and a set of auxiliary fields, with the complex scalar fields written as
\begin{equation}
    \Phi \: = \:
    \left(
    \begin{matrix}
        \phi \\
        \rho
    \end{matrix}
    \right)\,,
    \: \: \: \: \: \: \: \: \: \: \: \: \: \: \:
    \mathcal{F} \: = \:
    \left(
    \begin{array}{c}
        F \\
        G
    \end{array}
    \right)\,.
\end{equation}
Here, $F$ and $G$ are non-dynamical (non-propagating) in the sense that
their equations
of  motion are algebraic constraints that allow us to eliminate them from
the action. They have been introduced to ensure that the number of
bosonic and  fermionic degrees of freedom match off-shell and that the
algebra closes without use of the equations of motion.
After elimination of $F$ and $G$, the supersymmetry algebra 
no longer closes and
demonstration of invariance requires the use of
the equations of motion for the remaining fields.
The supersymmetry transformation parameter $\Theta$ forms a Dirac spinor
built from left- and right-handed Weyl parameters
\begin{equation}
    \Theta \: = \:
    \left(
    \begin{array}{c}
        \xi_{\alpha} \\
        \bar{\zeta}^{\dot{\alpha}}
    \end{array}
    \right)\,.
\end{equation}
Decomposing the Lagrangian \eqref{matter:Lagrangian}
into a sum of left- and right-handed parts yields
\begin{subequations} \label{Lagrangian:L:R}
  \begin{alignat}{1}
      \mathcal{L}_{left} \: & = \:
      ( \pmud \phi^{*} ) ( \pmuu \phi ) \: + \:
      i \: \bar{\psi}_{\dot{\alpha}}
      ( \bar{\sigma}^{\mu} )^{\dot{\alpha} \alpha} \pmud
      \psi_{\alpha} \: + \: F^{*}F\,, \nonumber \\
      \mathcal{L}_{right} \: & = \:
      ( \pmud \rho^{*} ) ( \pmuu \rho ) \: + \:
      i \: \chi^{\alpha}
      ( \sigma^{\mu} )_{\alpha \dot{\alpha}} \pmud
      \bar{\chi}^{\dot{\alpha}} \: + \: G^{*}G\,,
  \end{alignat}
\end{subequations}
which is a sum of two Wess-Zumino actions, one each for left-
and right-handed multiplets.

The Lagrangian \eqref{Lagrangian:L:R} is invariant,
up to a total divergence which does not
contribute to the action (at least for topologically trivial
field configurations), under the following set of supersymmetry
transformations, which we write in Weyl component form, forming
a closed algebra on the fields
\begin{eqnarray}
    \delta_{\xi} \psi_{\alpha}  & = &
    \sqrt{2} \left( i \: ( \sigma^{\mu} )_{\alpha \dot{\beta}}
    \bar{\xi}^{\dot{\beta}} \partial_{\mu} \phi
    \: + \: \xi_{\alpha} F  \right)\,,\nonumber \\
    \delta_{\zeta} \bar{\chi}^{\dot{\alpha}} & = &
    \sqrt{2} \left( i \: ( \bar{\sigma}^{\mu} )^{\dot{\alpha} \beta}
    \zeta_{\beta} \partial_{\mu} \rho^{*}
    \: + \: \bar{\zeta}^{\dot{\alpha}} G^{*} \right)\,,\nonumber \\
    \delta_{\xi} \phi & = & \sqrt{2} \: \xi^{\alpha} \psi_{\alpha}\,, \\
    \delta_{\zeta} \rho & = & \sqrt{2} \:
    \zeta^{\alpha} \chi_{\alpha}\,,\nonumber \\
    \delta_{\xi} F & = & i \sqrt{2} \: \bar{\xi}_{\dot{\alpha}}
    ( \bar{\sigma}^{\mu} )^{\dot{\alpha} \beta} \partial_{\mu}
    \psi_{\beta}\,,\nonumber \\
    \delta_{\zeta} G & = & i \sqrt{2} \:
    \bar{\zeta}_{\dot{\alpha}} ( \bar{\sigma}^{\mu} )^{\dot{\alpha} \beta}
    \partial_{\mu} \chi_{\beta}\,.\nonumber
\end{eqnarray}

We now take the fields in the left- and right-handed multiplets to
transform under the same supersymmetry transformations
(yielding $N=1$ supersymmetry), with $\Theta$ a Majorana
spinor, and apply the Noether procedure to derive the
coupling of the supergravity multiplet to the matter fields.
The gravitino will appear as the gauge field of local supersymmetry.
Supersymmetry breaking will manifest itself through the appearance of a mass for the gravitino.

\begin{figure}[htbp]
\centerline{
\psfig{figure=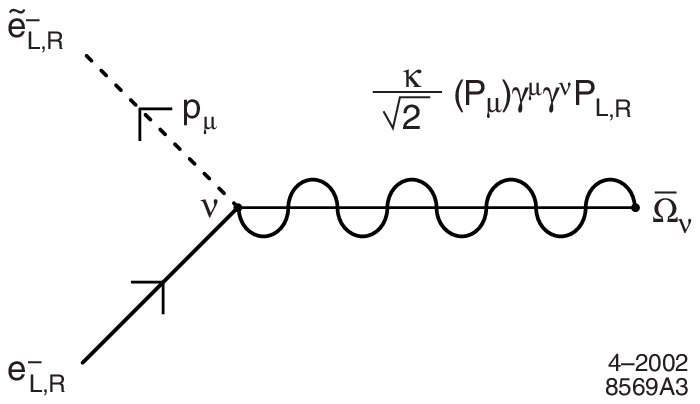,height=6cm,width=9cm,angle=0}}
\caption{Feynman rule for the gravitino, fermion, scalar coupling.
$P_{L,R}$ represent the standard projection operators.}
\label{8782_fig1}
\end{figure}

To gauge the global symmetry, we let the supersymmetry transformation parameter
$\Theta$ become space-time dependent $\Theta \rightarrow \Theta(x)$.
The Lagrangian will no longer be invariant, but will vary as
\begin{equation} \label{variation:1}
  \delta \mathcal{L} \: = \:
  J^{\mu} \partial_{\mu} \Theta \,,
\end{equation}
with $J^\mu$ the global supersymmetry current.
We must now add terms to the Lagrangian and the supersymmetry transformation
rules until we restore the invariance of the Lagrangian (up to a total
derivative).
We add a term coupling the local supersymmetry gauge field to the
symmetry current
\begin{equation}
  \mathcal{L}_1 \: = \: c \: \bar\Omega_\mu \: J^\mu\,,
\end{equation}
with $\bar\Omega_\mu$ a Majorana
vector-spinor, which we expect to become the gravitino,
and hence must transform as
\begin{equation}
  \delta \Omega_\mu \: = \: \frac{2}{\kappa} \partial_\mu \Theta
\end{equation}
to leading order in $\kappa$
in the supergravity theory (here $\kappa = \sqrt{8 \pi G_N}$).
The local variation vanishes to $O(\kappa^0)$ if $c=-\frac{\kappa}{2}$,
but not to order $\kappa$. Iterating this process to higher
orders in $\kappa$ and covariantizing with respect to gravity,
we arrive at a locally supersymmetric Lagrangian at order $\kappa^2$.  The
result consists of the
Einstein-Hilbert and the Rarita-Schwinger actions
together with the
original action \eqref{matter:Lagrangian} minimally coupled to gravity, plus a
term coupling spinors, scalars and gravitinos,
and higher order four point terms.
The term coupling scalars, spinors and gravitinos,
and minimally coupled to gravity, is
\begin{equation}
  \mathcal{L}_I \: = \:
  - \frac{\kappa}{\sqrt{2}} | e |
  \left\{
    \left( \partial_\mu \Phi_L \right) \bar{\Omega}_\nu
    \gamma^\mu \gamma^\nu \psi_L \: + \:
    \left( \partial_\mu \Phi_R \right) \bar{\Omega}_\nu
    \gamma^\mu \gamma^\nu \psi_R
  \right\}\, \: + \: h.c. \,,
\end{equation}
with $\bar{\Omega}_\nu$ a Majorana vector-spinor.
Expanding $|e|$ to leading order in the vierbein yields the Feynman rule
displayed in Fig. \ref{8782_fig1}, with the obvious generalization to
the hermitian conjugate piece.

\nn

\section{Phenomenological Analysis and Numerical Results}

We are now ready to apply our results to a phenomenological analysis.
For purposes of illustration we focus on selectron pair production
in high energy polarized \epem\
collisions.  As discussed in the introduction, this process provides
a valuable tool within the MSSM for determining the composition of the
mixed neutralino states, $\tilde\chi^0_i$ in terms of the various pure
SU(2) and U(1) wino and bino components, $\tilde W^0$, $\tilde B^0$.  Here,
we investigate how the existence of supersymmetric extra dimensions
modifies this reaction.

\begin{figure}[htbp]
\centerline{
\psfig{figure=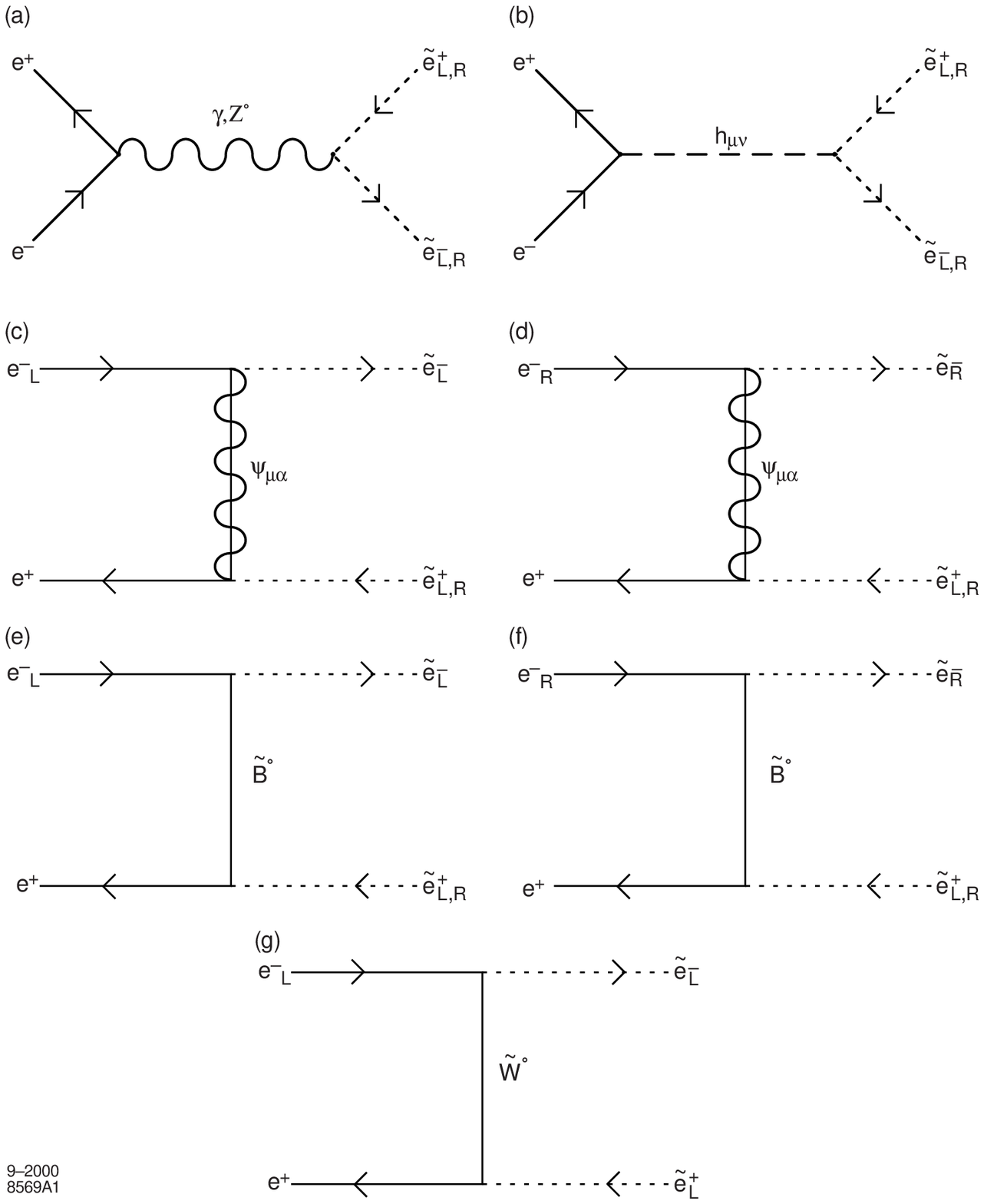,height=20cm,width=15cm,angle=0}}
\caption{Processes contributing to scalar electron pair production
in polarized \epem\ collisions.  The state $h_{\mu\nu}$ represents the
bulk graviton KK tower, $\Psi_{\mu\alpha}$ the bulk gravitino KK
tower, and $\tilde B^0,\, \tilde W^0$ correspond to the unmixed MSSM 
gaugino states.}
\label{8782_fig2}
\end{figure}

The tree level processes contributing to selectron pair
production in the presence of a supersymmetric bulk are
presented in Fig. \ref{8782_fig2}.  In addition to the standard
$\gamma,Z$ s-channel exchange and $\tilde B^0,\tilde W^0$ t-channel
contributions present in the MSSM, we now have contributions arising from
the s-channel exchange of the bulk graviton KK tower and the
t-channel exchange of the bulk gravitino KK tower.
There are no u-channel contributions due to the non-identical final
states.  The contributions from neutral higgsino states
are negligible due to the smallness of the Yukawa coupling.
The diagrammatic contributions to the individual
scattering processes for left- and right-handed selectron production with
initial polarized electron beams are summarized in Table \ref{diag}.
Note that the $\tilde W^0$ exchange only contributes to the process
$e^-_Le^+\to\sl^-\sl^+$, and that the t-channel gravitino and 
the $\tilde B^0$
contributions are isolated in the reaction 
$e^-_{L,R}e^+\to\tilde e^-_{L,R}\tilde e^+_{R,L}$.

\nn

\begin{table}
\centering
\begin{tabular}{|c|c|c|c|c|} \hline\hline
 & $\tilde e^-_L\tilde e^+_L$ & $\tilde e^-_R\tilde e^+_L$ &
   $\tilde e^-_L\tilde e^+_R $ & $\tilde e^-_R\tilde e^+_R$ \\ \hline
$e^-_Le^+$ & s-channel $\gamma\,, Z\,, G_n$ & & & s-channel $\gamma\,, Z\,,
  G_n$ \\
  & t-channel $\tilde W\,, \tilde B\,, \Psi_n$ & & t-channel $\tilde B\,,
    \Psi_n$ & \\ \hline
$e^-_Re^+$ & s-channel $\gamma\,, Z\,, G_n$ & & & s-channel $\gamma\,, Z\,,
G_n$ \\
  & & t-channel $\tilde B\,, \Psi_n$ & & t-channel $\tilde B\,, \Psi_n$ \\
\hline\hline
\end{tabular}
\caption{The diagrammatic contributions to individual scattering processes
for polarized electron beams.  A blank box indicates that there are no
contributions for that polarization configuration.}
\label{diag}
\end{table}

Our amplitudes for the standard MSSM contributions to the reaction
$e^-(p_2)e^+(p_1)\to\tilde e^-(k_2)\tilde e^+(k_1)$ (with the direction
of the charge flow as indicated in Fig. \ref{8782_fig2}) reproduce the
results in \cite{Baer:1988kx}.
The unpolarized matrix element for the case of massive gravitino KK
exchanges is
\begin{equation}
{\cal M}={\frac{\kappa^2}{2}}\sum_{\vec n} {\frac{k_1^\nu k_2^\rho}
{t-m^2_{\vec n}}}\bar e(p_1)\gamma_\mu\gamma_\nu P^{\vec n,\mu\tau}
\gamma_\rho\gamma_\tau e(p_2)\,,
\end{equation}
where the sum extends over the gravitino KK modes and we recall that
$\kappa=\sqrt{8\pi G_N} =\bar M_P^{-1}$
is the reduced Planck scale.   $P^{\vec n,\mu\tau}$ represents the
numerator of the propagator for a Rarita-Schwinger field of mass
$m_{\vec n}$ and is given in Appendix \ref{Propagator:sum}.
The mass splitting between the evenly spaced bulk gravitino KK
excitations is given by
$1/R_c$, which lies in the range $10^{-4}$ eV to few MeV for $\delta=2$ to
6 assuming $M_D\sim 1$ TeV; their number density is thus large at
collider energies.
The sum over the KK states can then be approximated
by an integral
which is log divergent for $\delta=2$ and power divergent for
$\delta>2$.  We employ a cut-off to regulate these ultraviolet
divergences, with the cut-off being set to $\Lambda_c$, which in
general is different from $M_D$, to
account for the uncertainties from the unknown ultraviolet physics.
This approach is the most model independent and is that
generally used in the case of virtual graviton exchange \cite{Hewett:1998sn}.
In practice,
the integral over the gravitino KK states is more
complicated than that in the case with spin-2 gravitons due to the 
dependence of the gravitino propagator on $m_{\vec n}$.
We find that the leading order term for $\sqrt{|t|}\ll\Lambda_c$
results in
the replacement (in the case of $\delta=6$)
\begin{equation}
{\frac{\kappa^2}{2}}\sum_{\vec n} {\frac{P^{\vec n,\mu\tau}}{t-m^2_{\vec n}}}
\to {\frac{-i8\pi}{5\Lambda_c^3}}\left( \eta^{\mu\nu}-{\frac{1}{3}}
\gamma^\mu\gamma^\nu\right) \,
\end{equation}
in the matrix element; the structure of the summed gravitino propagator
is thus altered from that of a single massive state.
Hence the leading order behavior for gravitino KK exchange results in
a dimension-6 operator!  This is in stark contrast to graviton KK
exchange which yields a dimension-8 operator at leading order.
We thus expect an increased sensitivity to the fundamental scale $M_D$ in
the case of a supersymmetric bulk.  Our derivation of
the summation over the bulk gravitino KK states in the matrix element
is detailed in Appendix C.
Lastly, the bulk graviton KK tower contributes an additional
amplitude of the form \cite{tgr}
\begin{equation}
{\cal M} = {\frac{2}{\Lambda_c^4}}(t-u)(k_1-k_2)_\mu\bar e(p_2)
\gamma^\mu e(p_1)\,.
\end{equation}

In order to perform a numerical analysis of this process, we need to specify
a concrete supersymmetric model.  We choose that of Gauge 
Mediated Supersymmetry
Breaking (GSMB) as it naturally contains a light zero-mode gravitino.
We specify a sample set of input parameters at the messenger scale, where the
supersymmetry breaking is mediated via the messenger sector, and
use the Renormalization Group Equations (RGE) to obtain the low-energy
sparticle spectrum.  We choose two sets of sample input parameters
describing the messenger sector which are consistent with our model.
The RGE evolution of these parameter sets is performed in ISAJET v7.51
\cite{frank} and results in the sparticle spectrum
\begin{eqnarray}
{\rm Set~I} : &  m_{\tilde e_L} & =217.0\, {\rm GeV},\quad
m_{\tilde e_R}=108.0\, {\rm GeV}, \quad
\chi^0_i=(76.5,\, 141.5,\, 337.0,\, 367.0)\, {\rm GeV} \,,\nonumber \\
{\rm Set~II}: &  m_{\tilde e_L} & =210.5\, {\rm GeV},\quad
m_{\tilde e_R}=104.5\, {\rm GeV}, \quad
\chi^0_i=(110.5,\, 209.6,\, 322.5,\, 324.0)\, {\rm GeV}  \,,
\nonumber
\end{eqnarray}
where $\chi^0_i$ with $i=1,4$ corresponds to the four mixed neutralino
states.
The first set of parameters yields a bino-like state for the lightest
neutralino, whereas
the second set results in a Higgsino-like state for $\chi^0_1$.
These two sets are chosen so we can investigate the dependence of the
kinematic distributions on the composition of the lightest neutralino
state.  Note that the \sl\ and \sr\ masses are essentially equivalent
between the two sets and hence will not induce any kinematical
differences in the distributions.  In addition,
these input parameters were selected in order to obtain a sparticle spectrum
which is kinematically accessible to the Linear Collider; our
results are essentially insensitive to the exact details of the spectrum.
We stress that our analysis is purely phenomenological and that our
conclusions do not depend on the physics inherent to GMSB.
Except where noted, we perform our numerical analysis for the case
of $\delta=6$, following our above discussion of supergravity models.
From here on, we refer to these two spectra as our $D=4$ supersymmetric
models.

\begin{figure}[htbp]
\centerline{
\psfig{figure=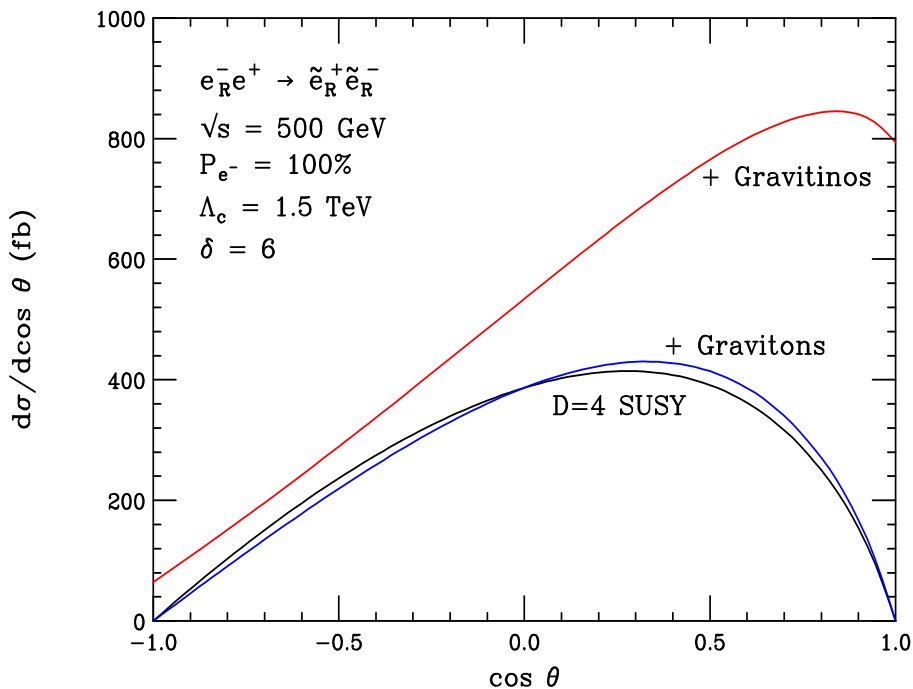,height=9cm,width=12cm,angle=90}}
\caption{The angular distribution for $e^-_Re^+\to\sr^+\sr^-$ from
the $D=4$ supersymmetric model I, plus the addition of bulk graviton
KK tower exchange, and with bulk gravitino KK tower exchange,
corresponding to the bottom, middle, and top curves, respectively.}
\label{8782_fig3}
\end{figure}

It is instructive to first examine the effects of each class of contributions
to selectron pair production.  This is displayed
in Fig \ref{8782_fig3}, which shows the angular distribution for
the process $e^-_Re^+\to\sr^+\sr^-$ with $\sqrt s=500$ GeV assuming
100\% polarization of the electron beam; we show
this particular reaction merely for purposes of demonstration.
The bottom curve represents the full contributions
(s- and t-channel) from the 4-dimensional standard
gauge-mediated supersymmetric model discussed above in the case where 
the $\chi^0_1$ is bino-like, corresponding to parameter set I.
Our numerical results for the MSSM case agree with those in the literature
\cite{Baer:1988kx}.  The middle curve displays the effects of adding only 
the s-channel contributions of the bulk graviton KK tower
in the scenario of a non-supersymmetric bulk with $\Lambda_c=1.5$ TeV.
We see that there
is little difference in the distribution between the $D=4$ supersymmetric
case and with the addition of the
graviton KK tower, in either shape or magnitude.  It would hence be 
difficult to disentangle the effects of graviton exchange from an accurate
measurement of the underlying supersymmetric parameters using this
process alone.  The top curve corresponds to the full set of contributions
from a supersymmetric bulk, \ie, our standard supersymmetric model plus
KK graviton and KK gravitino tower exchange
for the case of six extra dimensions with
$\Lambda_c=1.5$ TeV.  Here we see that the exchange of bulk gravitino 
KK states yields a large enhancement in the cross section and a substantial
shift in the shape of the angular distribution, particularly at forward
angles, even for $\Lambda_c=3\sqrt s$.
This provides a dramatic signal for a supersymmetric bulk!

\begin{figure}[htbp]
\centerline{
\psfig{figure=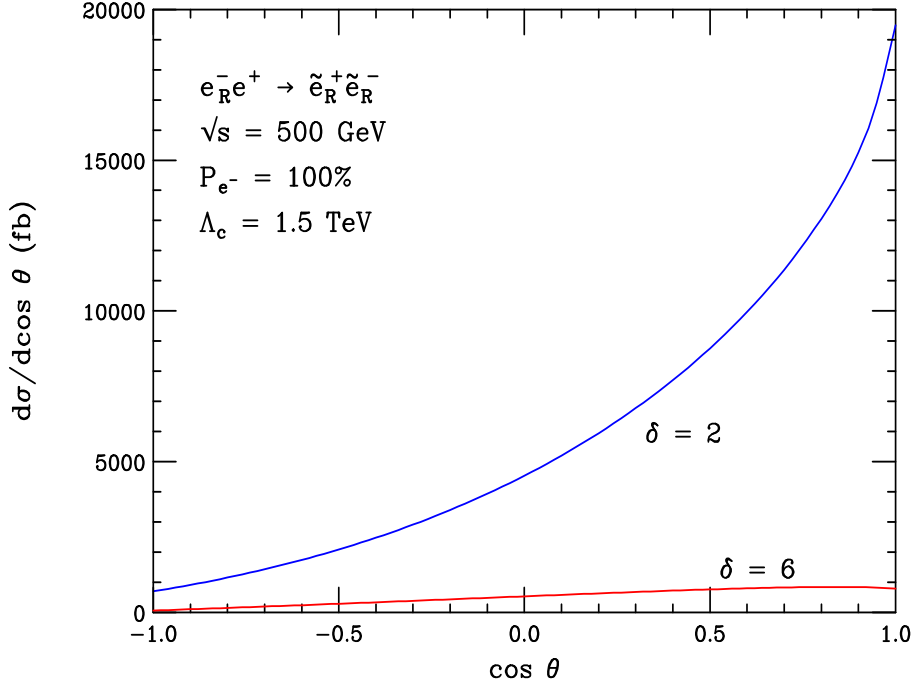,height=9cm,width=12cm,angle=90}}
\caption{A comparison of the event rates for the cases with
$\delta=2$ and 6,  corresponding
to the top and bottom curves, respectively.}
\label{8782_fig4}
\end{figure}

In Fig. \ref{8782_fig4} we explore the modifications to the 
angular distribution for $e^-_Re^+\to\sr^+\sr^-$
from a supersymmetric bulk as the value of $\delta$ changes,
using our parameter set I for demonstration.  In principle there are two 
competing effects which may modify the distributions:  (i) the number of
degenerate gravitinos in each KK level as a result of the reduction
of fermions,
versus (ii) the volume factor that appears in the density of states
in the integral over the gravitino
propagator which sums over the states in the KK tower, and its dependence
on $\Lambda_c$.
As discussed above, the number of degenerate gravitinos
is reduced as the number of extra dimensions decreases; this results
in a reduction of the cross section in the general case of extended low-energy
supersymmetries.  However, this effect does not modify our analysis since we
have assumed the $N=1$ at low-energy.  The second effect arises
from the increase in $R_c$, with $\Lambda_c$ being held fixed, 
for smaller values of $\delta$ as can be seen in Appendix C.
This volume factor arises in the
integral over the propagators for the bulk gravitino KK tower and is 
discussed in Appendix C.  In this figure, we compare the event rates for 
this process for the cases $\delta=2$ and 6, corresponding to the
top and bottom curves,  respectively.  We note that the shape of the
distribution differs in the two cases due to the form of the sub-leading
terms in the integral over the propagator of the KK states.
For the remainder of our analysis,  we will
display results only for the more conservative case of $\delta=6$.

\nn

\nn

\begin{figure}[htbp]
\centerline{
\psfig{figure=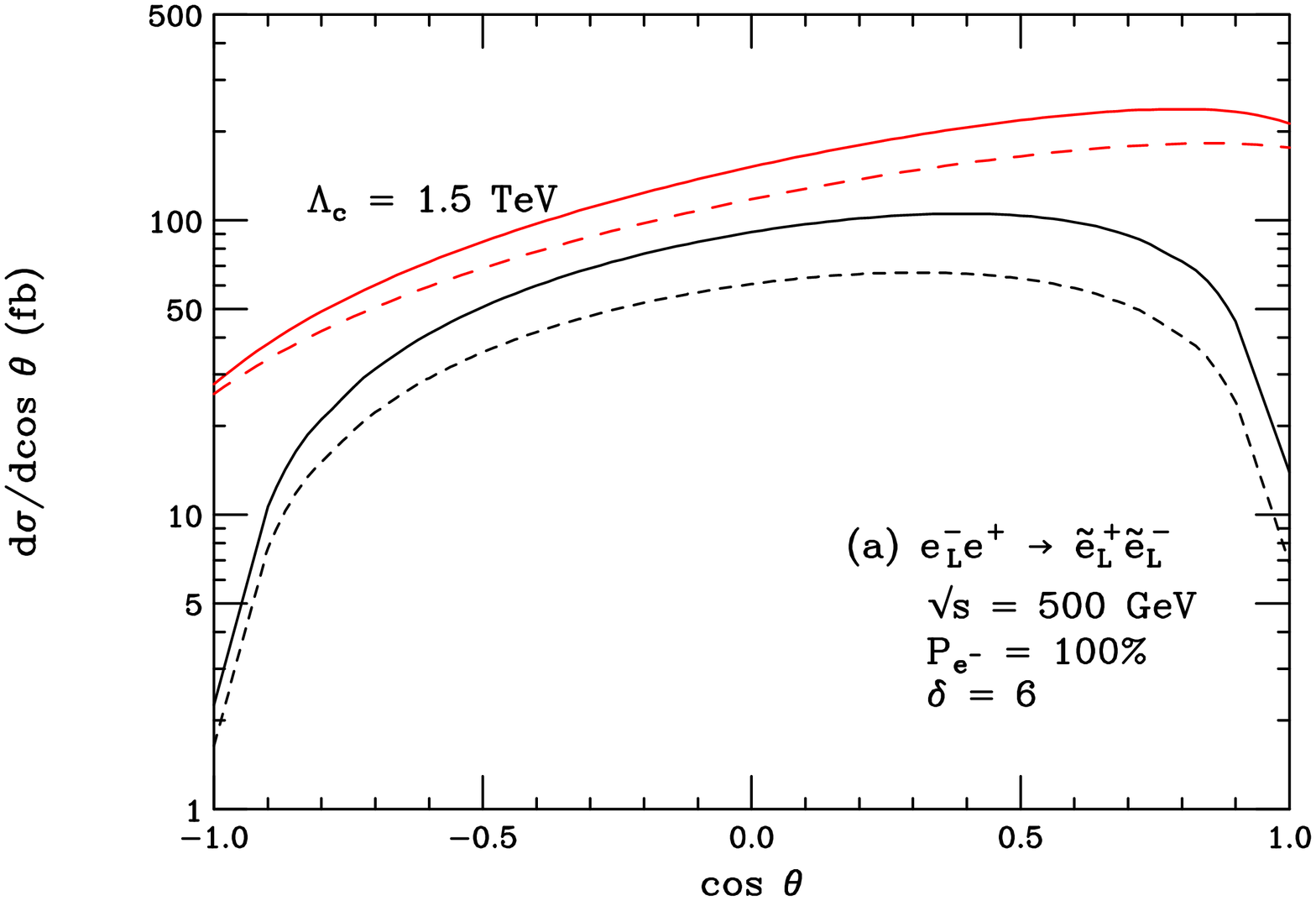,height=6.5cm,width=8cm,angle=90}
\hspace*{5mm}
\psfig{figure=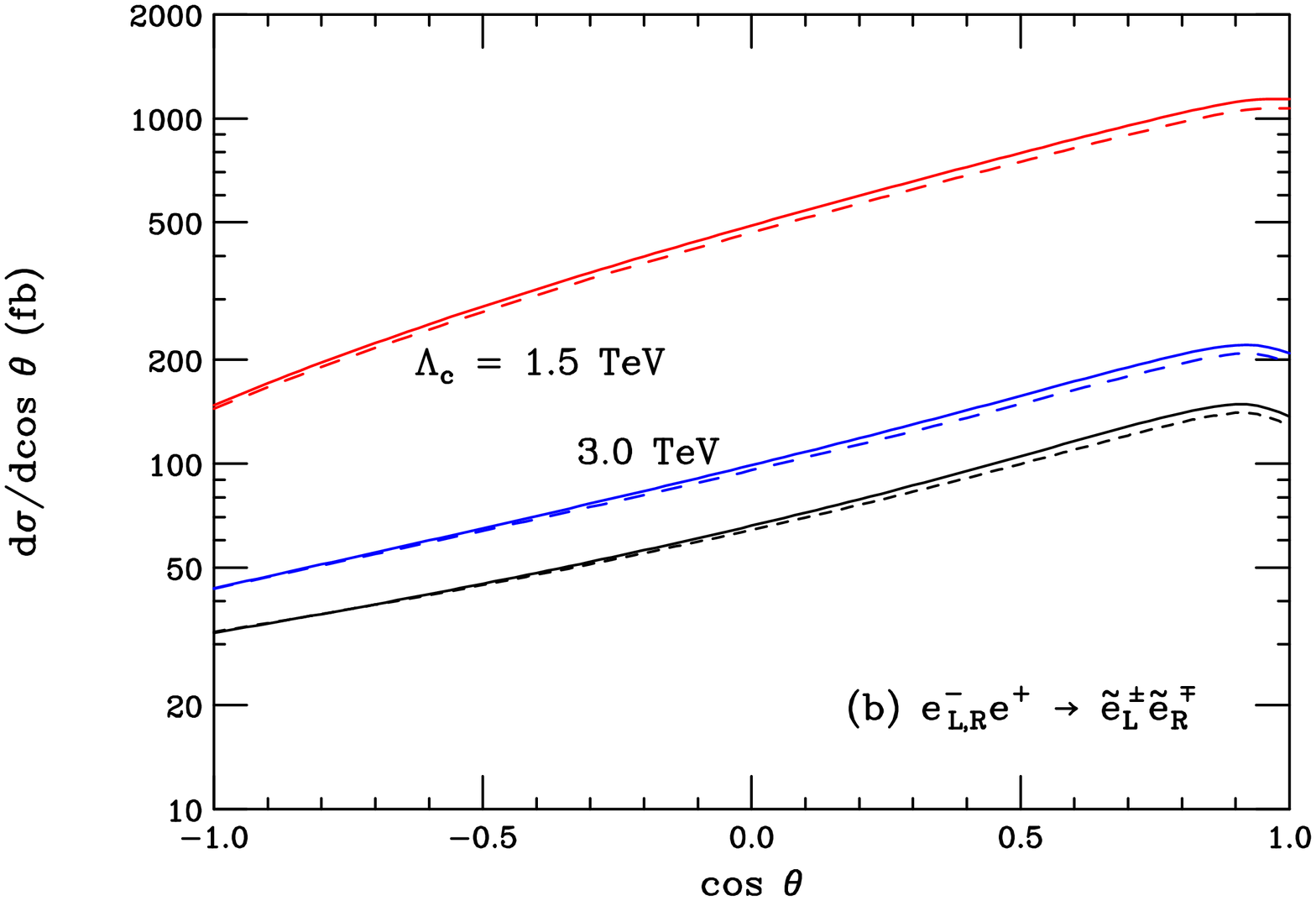,height=6.5cm,width=8cm,angle=90}}
\vspace*{0.25cm}
\centerline{
\psfig{figure=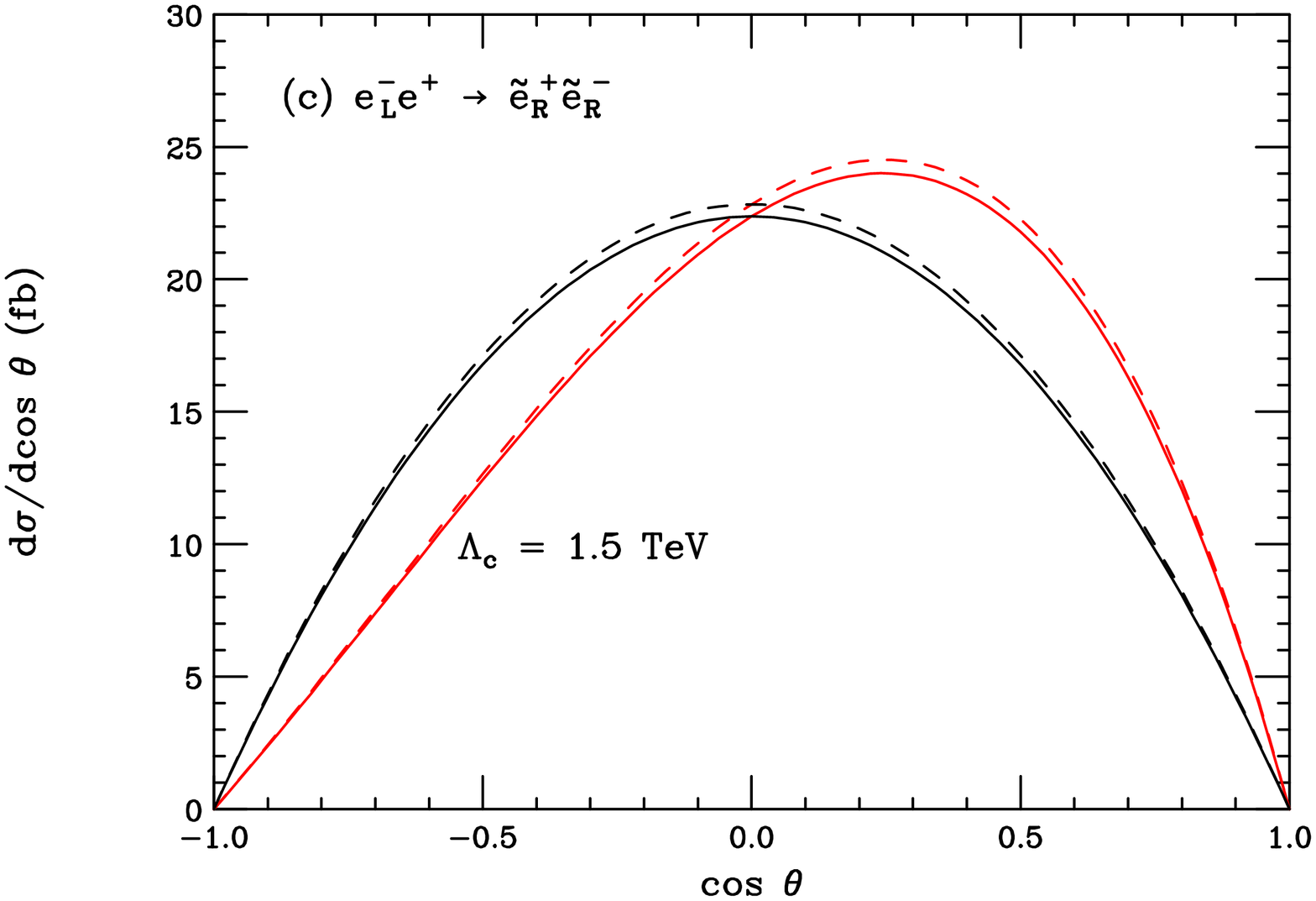,height=6.5cm,width=8cm,angle=90}
\hspace*{5mm}
\psfig{figure=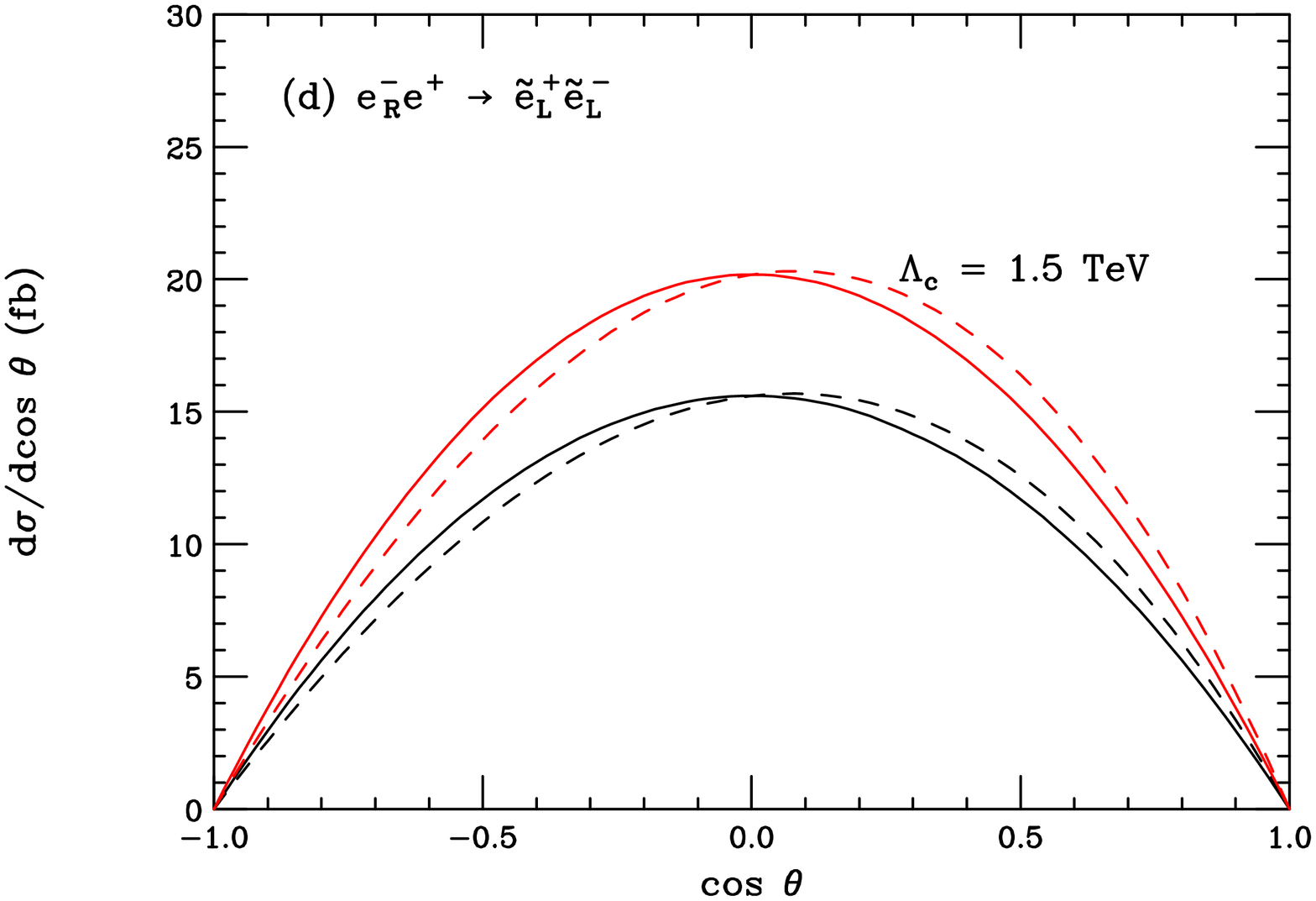,height=6.5cm,width=8cm,angle=90}}
\vspace*{0.25cm}
\centerline{
\psfig{figure=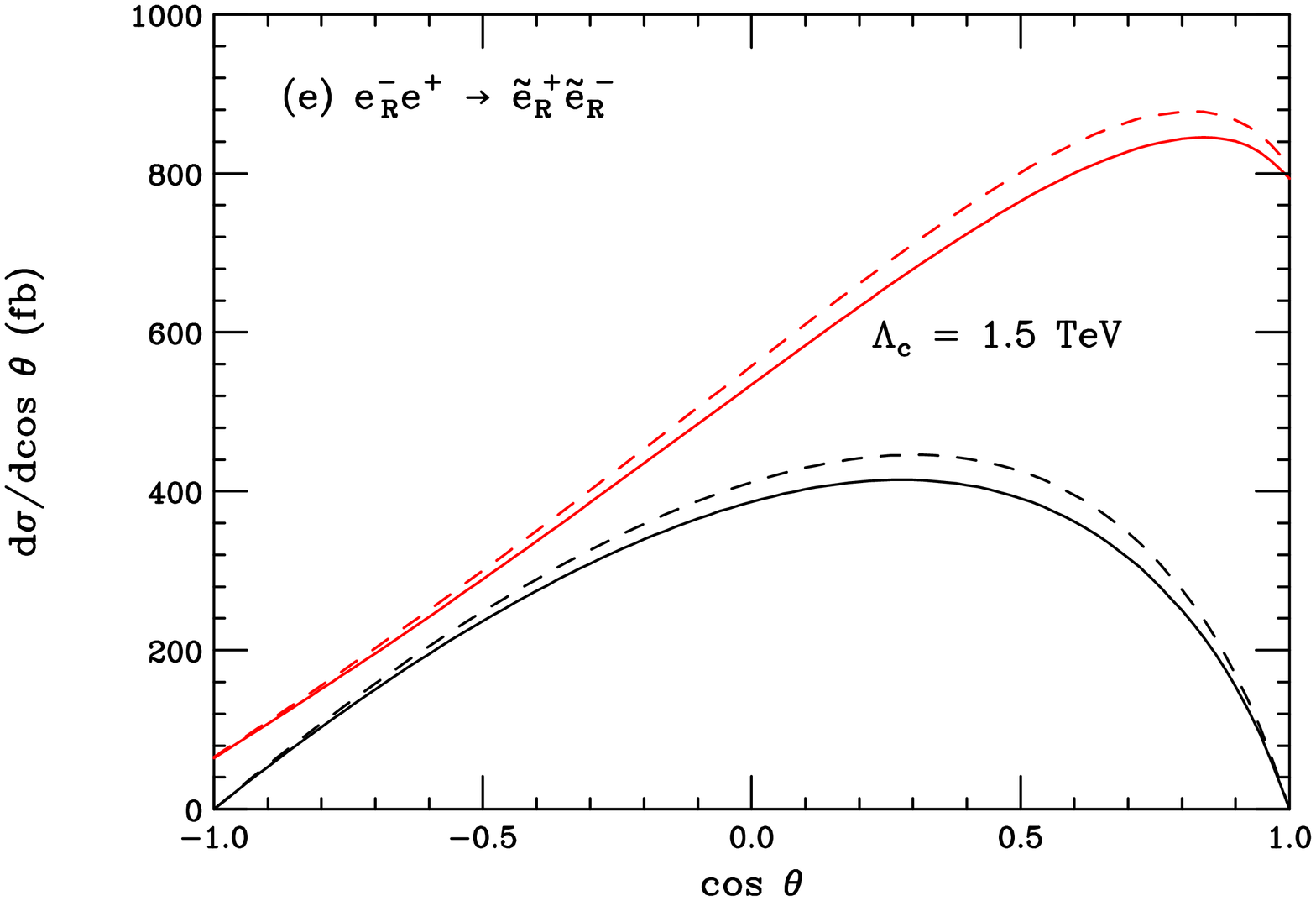,height=6.5cm,width=8cm,angle=90}}
\caption{Angular distributions for each helicity configuration with
supersymmetric bulk contributions for $\Lambda_c=1.5$ TeV (top
curves), and for  the $D=4$ supersymmetric models (bottom curves).
The solid (dashed) curves correspond to a bino-like (Higgsino-like)
composition of the lightest neutralino.}
\label{8782_fig5}
\end{figure}

Let us now study the variations in the distributions between the
two different compositions of the lightest neutralino.
Figure \ref{8782_fig5} shows the
angular distributions with $100\%$ electron beam polarization
for each helicity configuration listed in Table \ref{diag} for the two
sets of parameters discussed above, with and without the contributions
from supersymmetric extra dimensions.  In each case, the solid curve 
corresponds to the bino-like case and the dashed curve
represents the Higgsino-like scenario.  The top set of curves are those
for a supersymmetric bulk with $\Lambda_c=1.5$ TeV, while the
bottom set corresponds to our two $D=4$ supersymmetric models,
\ie, without the graviton and gravitino KK contributions.  
We note that the $D=4$ results agree with  those in the
literature \cite{Baer:1988kx}.  We see from the figure that in the 
process where the gravitino contributions are dominant, 
$e^-_{L,R}e^+\to\sl^\pm\sr^\mp$, there is little
difference in the shape or magnitude between the two $\chi^0_1$
compositions.  The use of
selectron pair production in polarized \epem\ collisions as a means of
determining the composition of the lightest neutralino is thus made
more difficult in the scenario with supersymmetric large extra dimensions.
In what follows, we present results only
for the bino-like $\chi^0_1$ as a sample case; our conclusions will not
be dependent on the assumptions of the composition of the lightest
neutralino.

\nn

\begin{figure}[htbp]
\centerline{
\psfig{figure=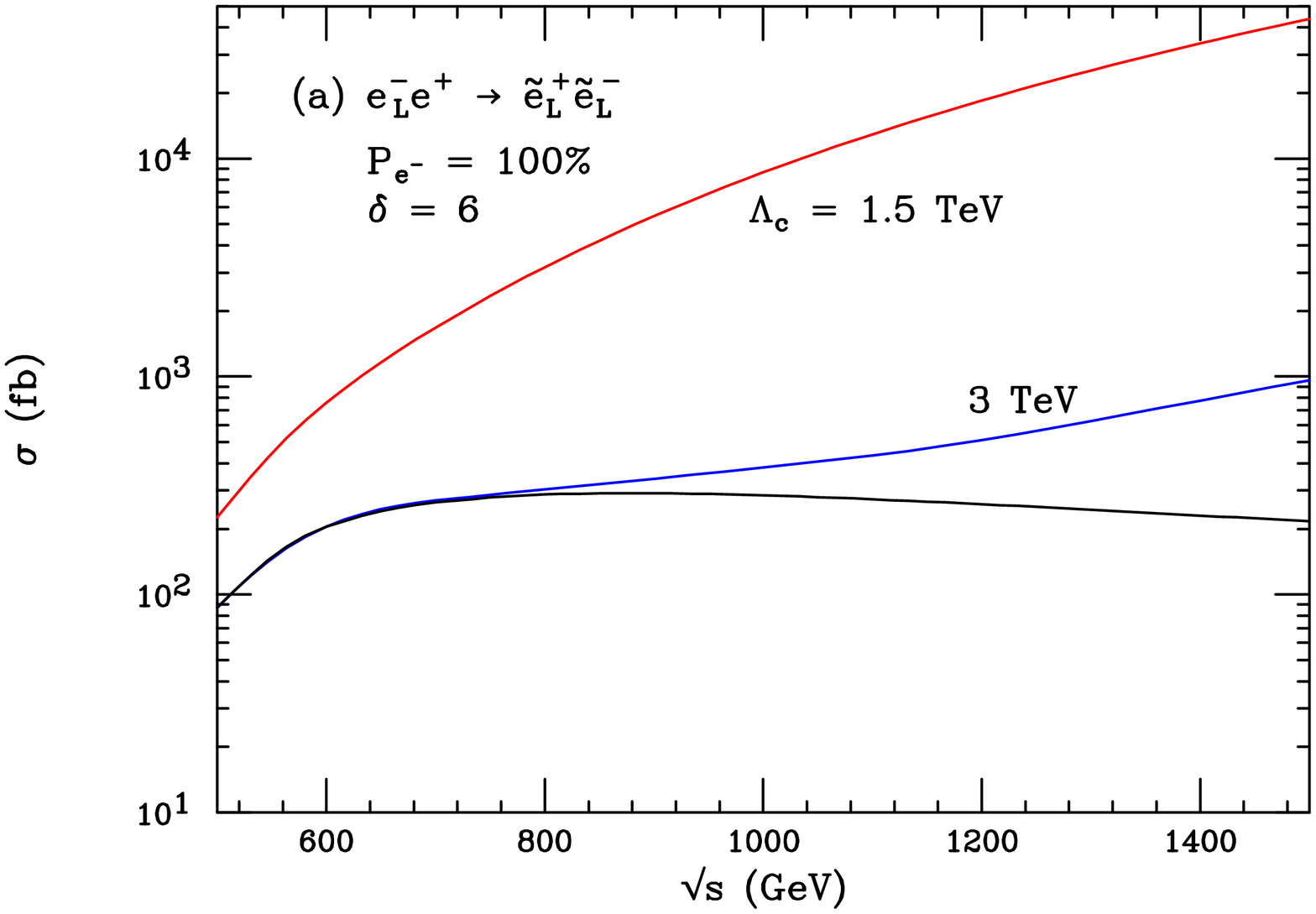,height=6.5cm,width=8cm,angle=90}
\hspace*{5mm}
\psfig{figure=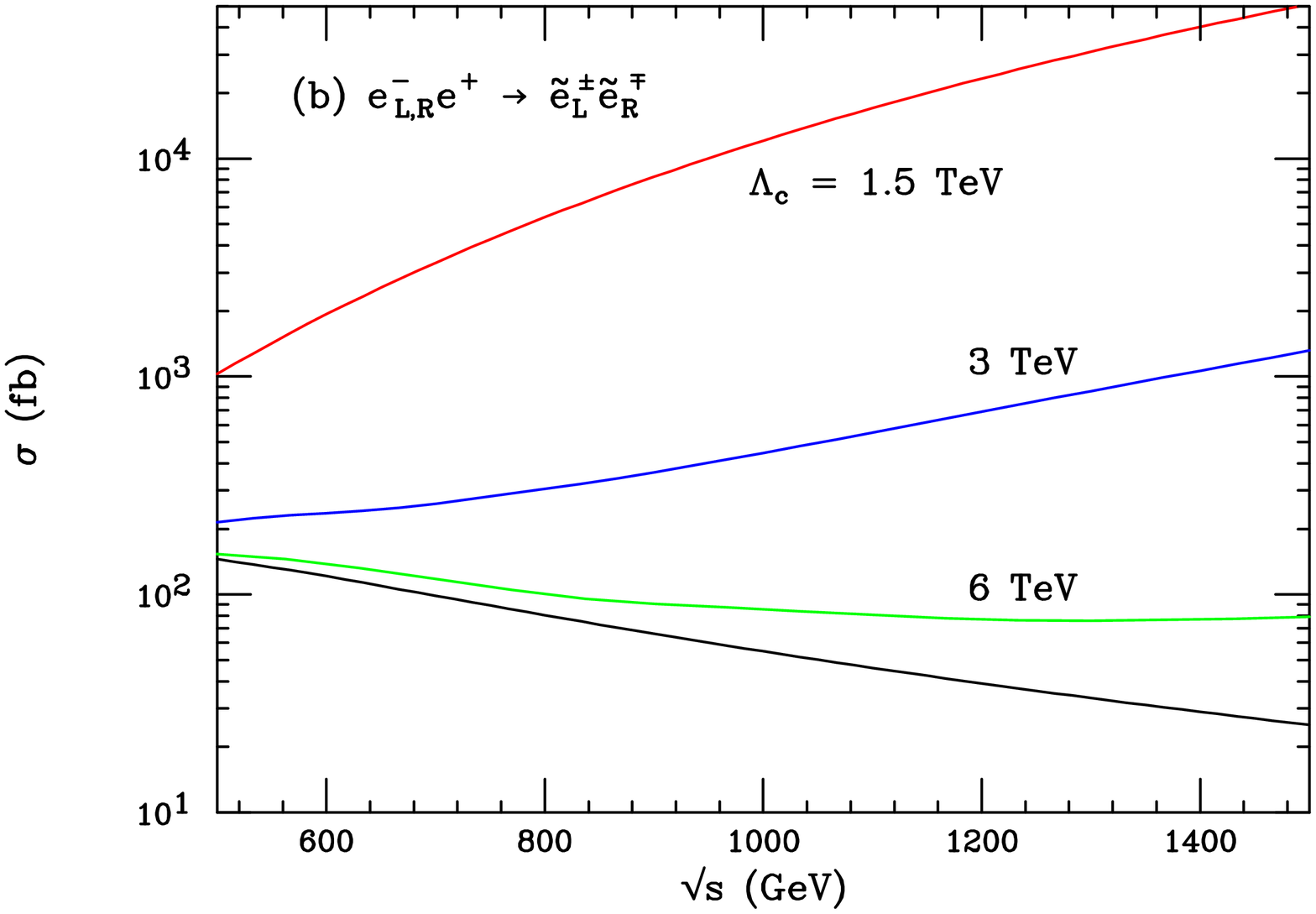,height=6.5cm,width=8cm,angle=90}}
\vspace*{0.25cm}
\centerline{
\psfig{figure=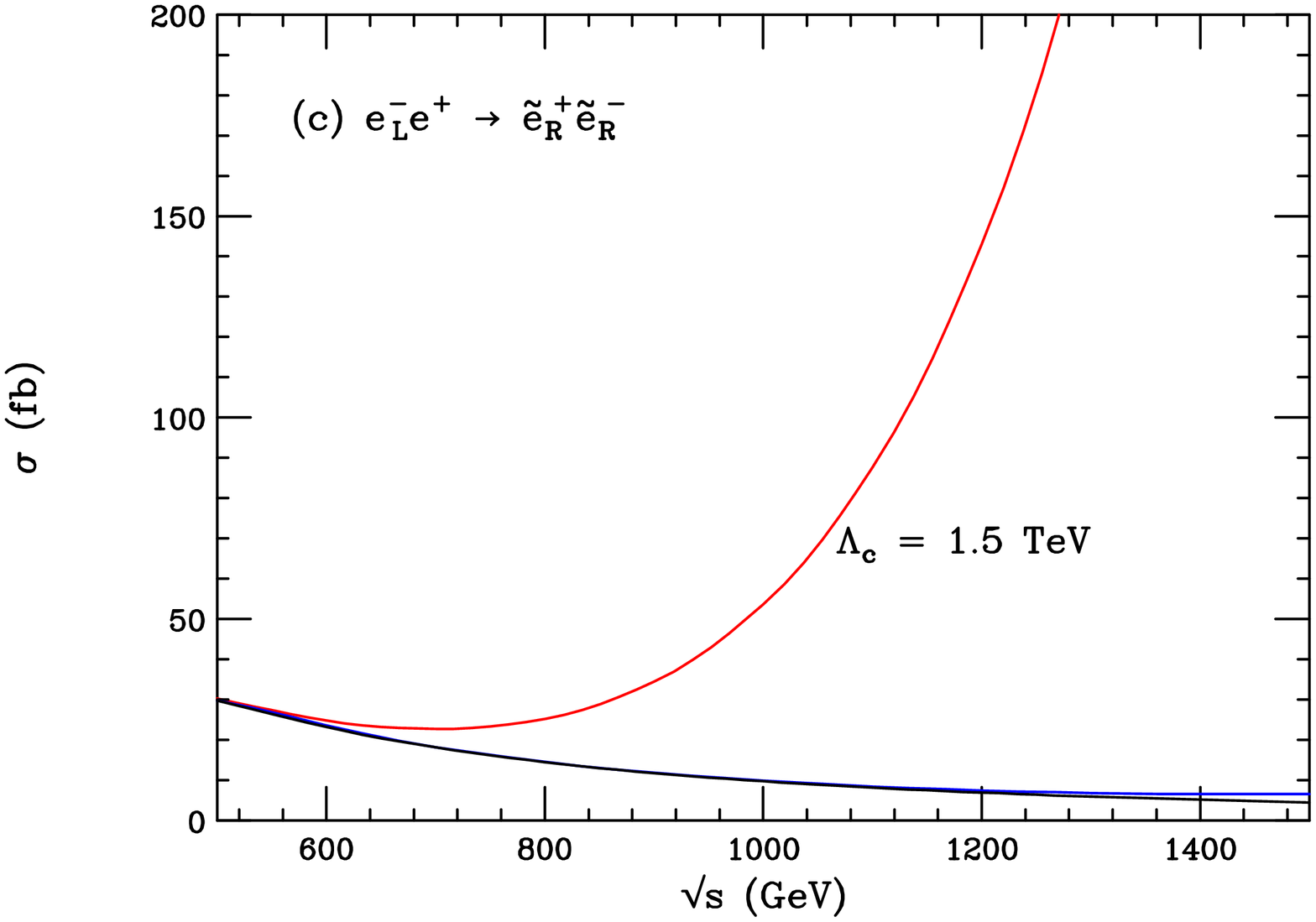,height=6.5cm,width=8cm,angle=90}
\hspace*{5mm}
\psfig{figure=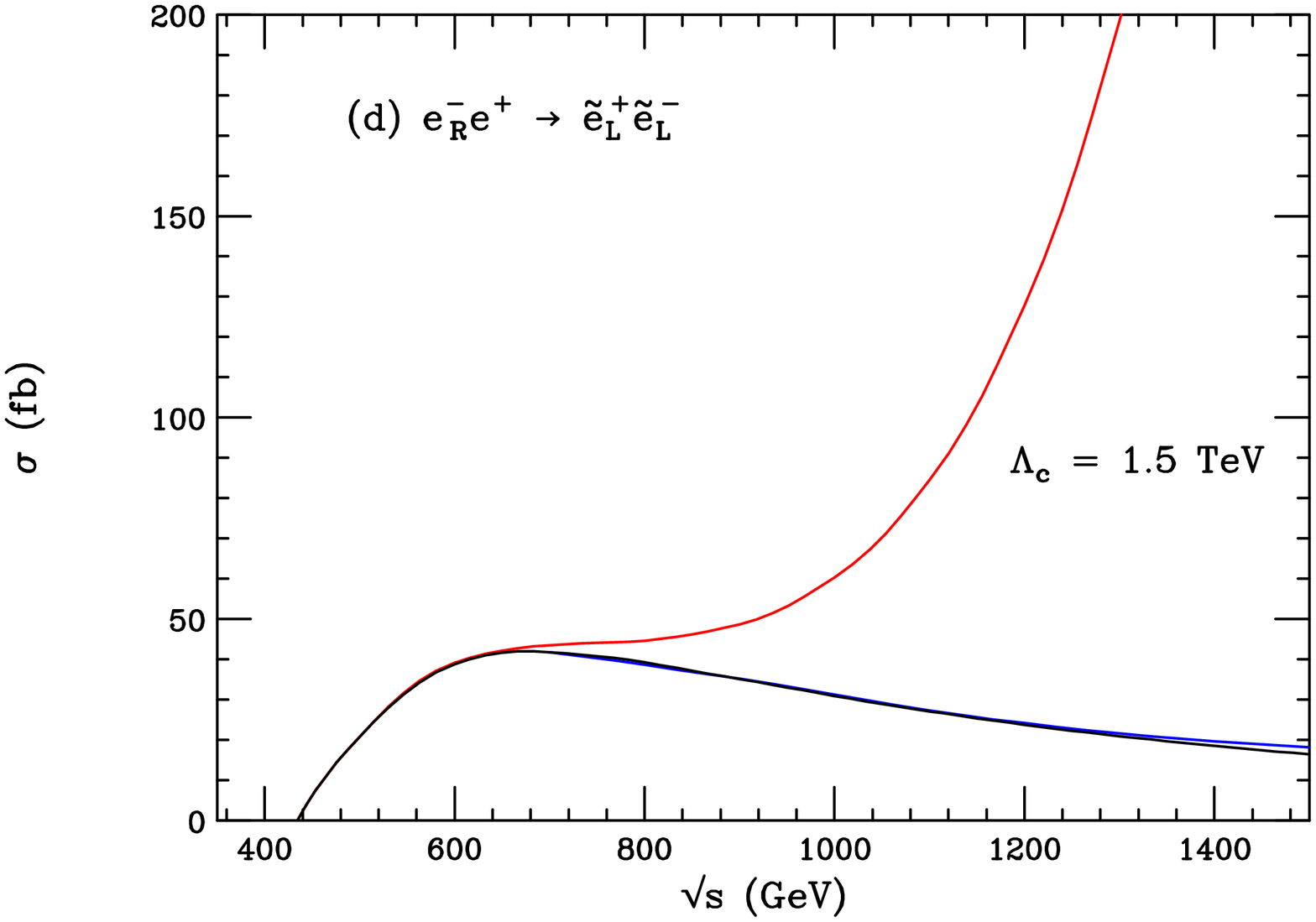,height=6.5cm,width=8cm,angle=90}}
\vspace*{0.25cm}
\centerline{
\psfig{figure=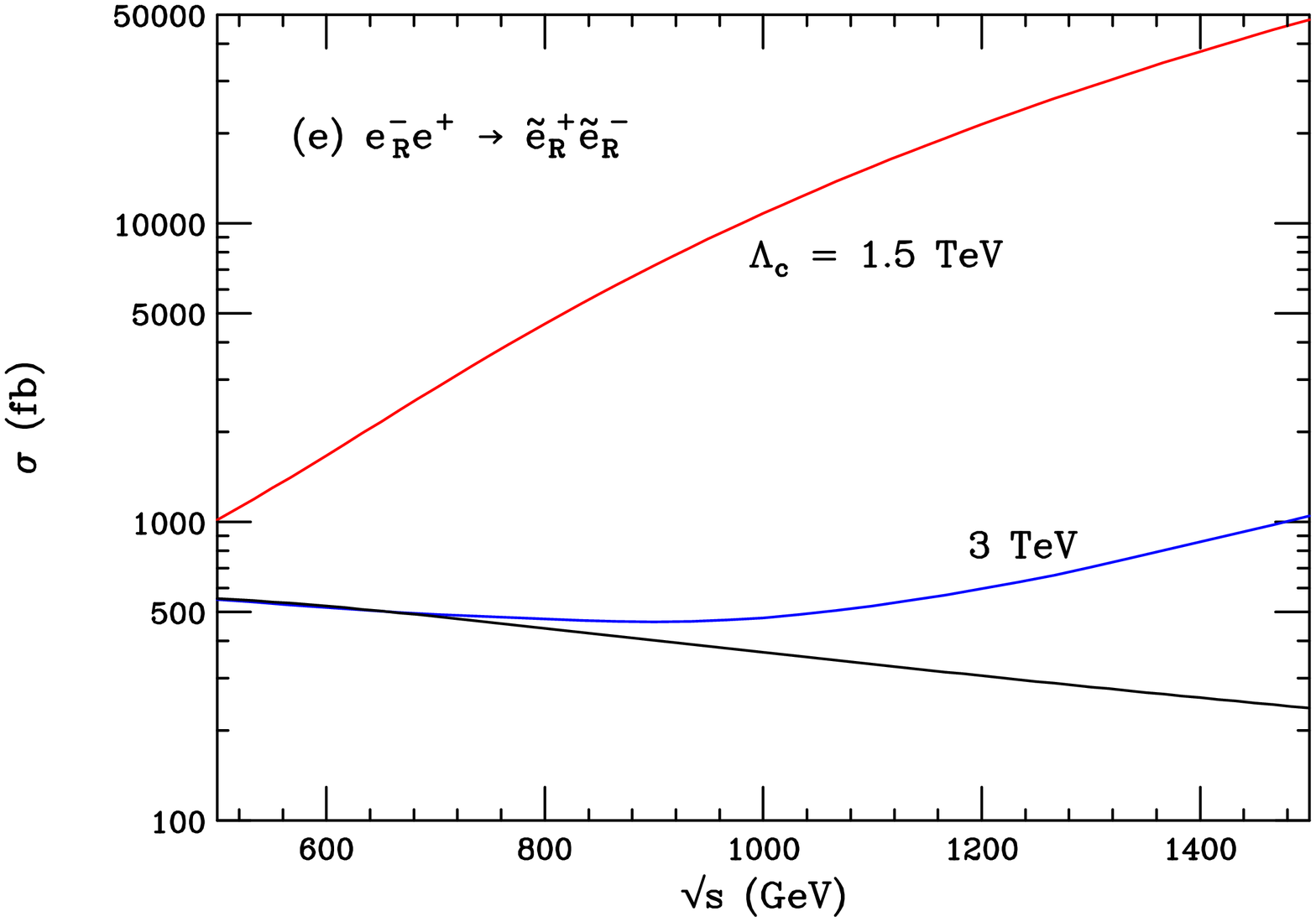,height=6.5cm,width=8cm,angle=90}}
\caption{Total cross sections as a function of center of mass energy.
The bottom curve corresponds to the $D=4$ supersymmetric model, and
the other curves are for a $\delta=6$ supersymmetric bulk with the
cut-off as labeled.  In figures (c) and (d), the $\Lambda_c=3$ TeV
curve is barely distinguishable from that for the case of 
$D=4$ supersymmetry.}
\label{8782_fig6}
\end{figure}

We now examine in Fig. \ref{8782_fig6} the total cross section as a 
function of center-of-mass energy for each helicity configuration.  
In each case, the bottom curve represents the $D=4$ bino-like 
supersymmetric model, while the remaining curves, from top to bottom,
are for a supersymmetric bulk with $\Lambda_c=1.5, 3.0,
6.0$ TeV.  In some reactions, the results for $\Lambda_c=6.0$ TeV are
indistinguishable from the $D=4$ case.
Here we can see the effects of unitarity violation as
$\sqrt s$ approaches the value of the cut-off scale.  Clearly, the new,
as of yet unknown, ultra-violet physics will set-in at this point to
regularize the cross section.

\nn

\begin{figure}[htbp]
\centerline{
\psfig{figure=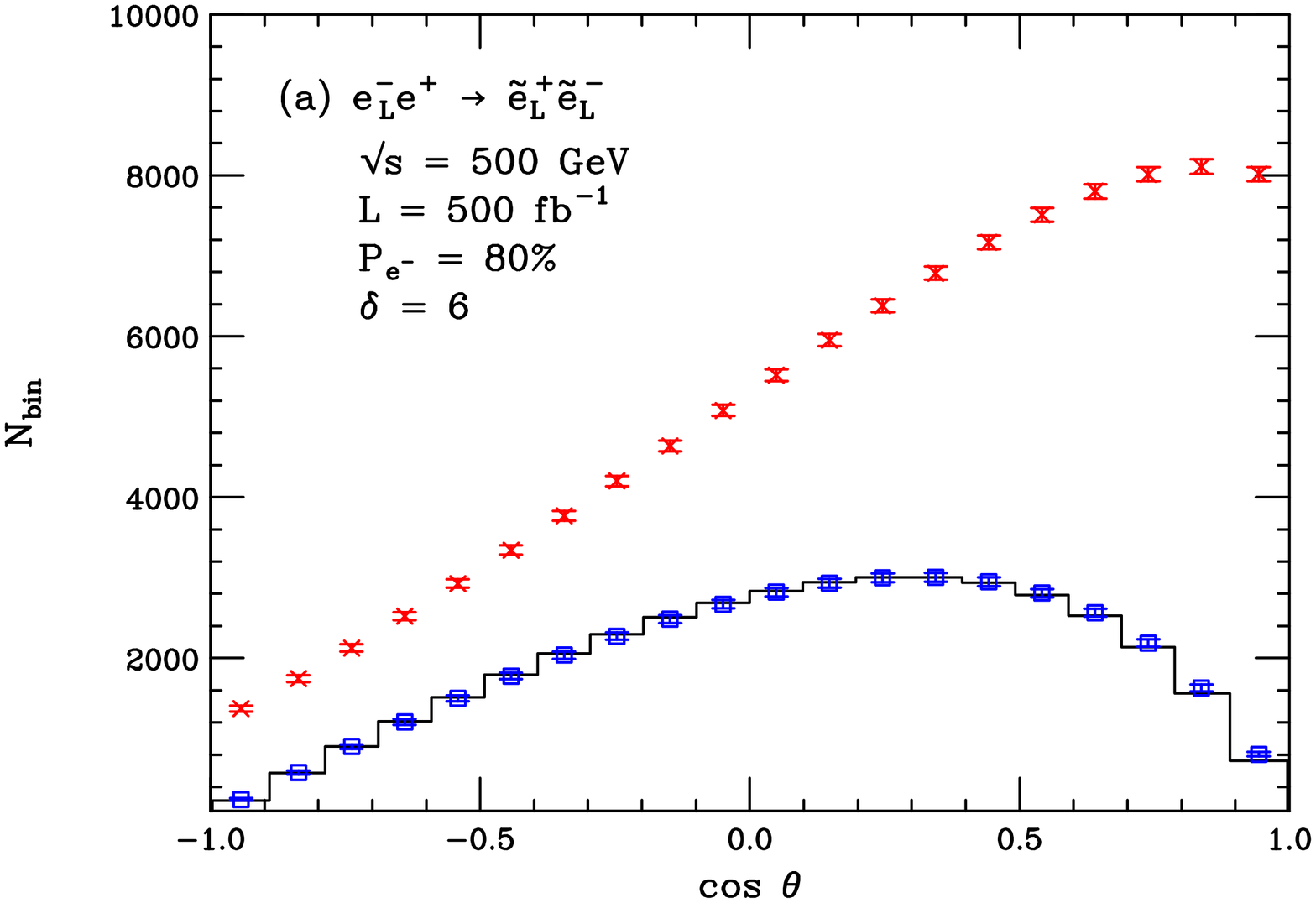,height=9cm,width=12cm,angle=90}}
\vspace*{0.25cm}
\centerline{
\psfig{figure=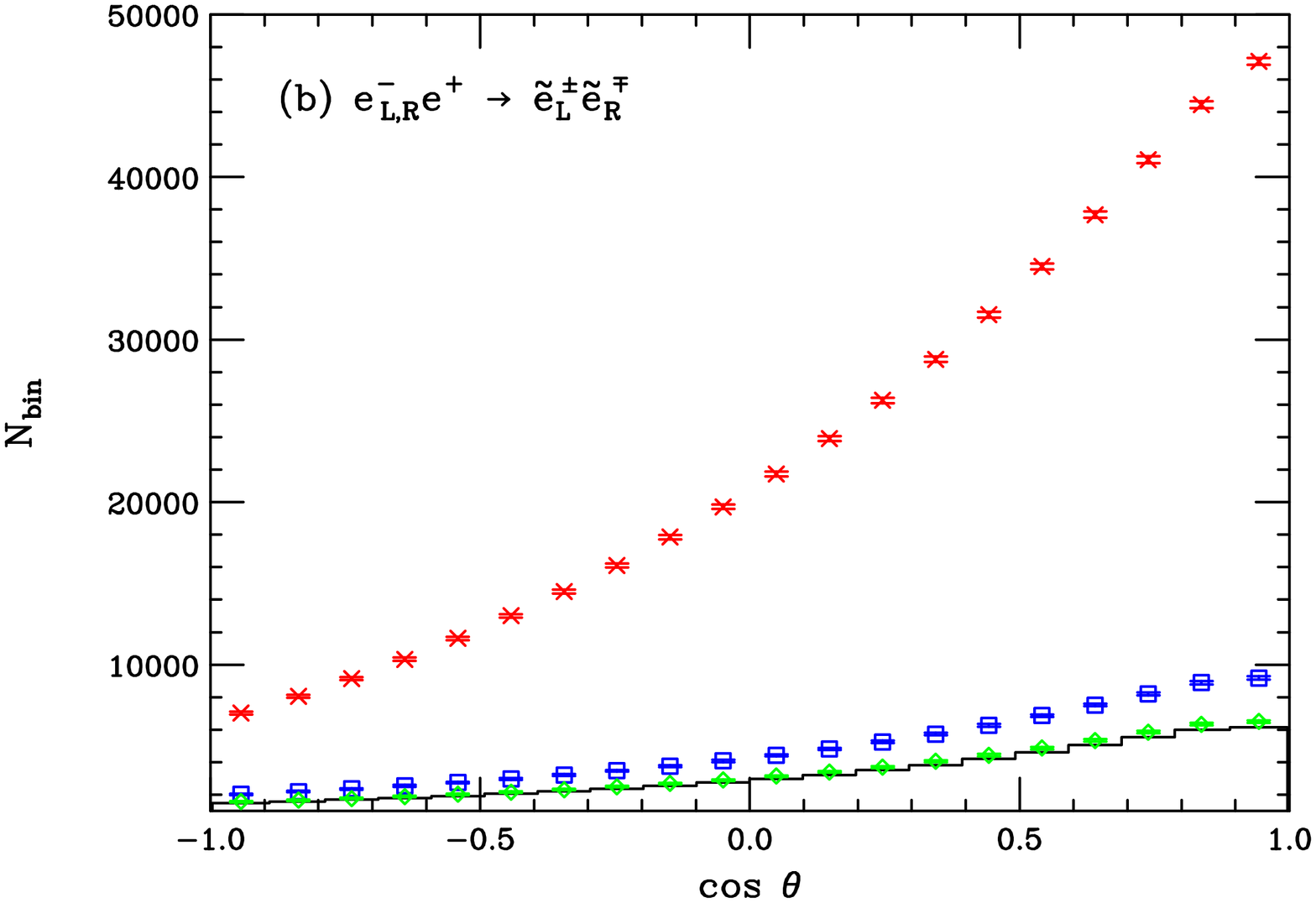,height=9cm,width=12cm,angle=90}}
\caption{(a-b) Polarized binned angular distributions for each helicity
configuration, taking an 80\% polarization of the initial electron
beam.  The solid histogram represents the $D=4$ bino-like model,
while the ``data'' points correspond to the effects of a supersymmetric
bulk with $\Lambda_c=1.5, 3.0, 6.0$ TeV from top to bottom.  The
$\Lambda_c=6.0$ TeV case is only discernable in Figure (b).}
\label{8782_fig71}
\end{figure}

\begin{figure}[htbp]
\vspace*{0.25cm}
\centerline{
\psfig{figure=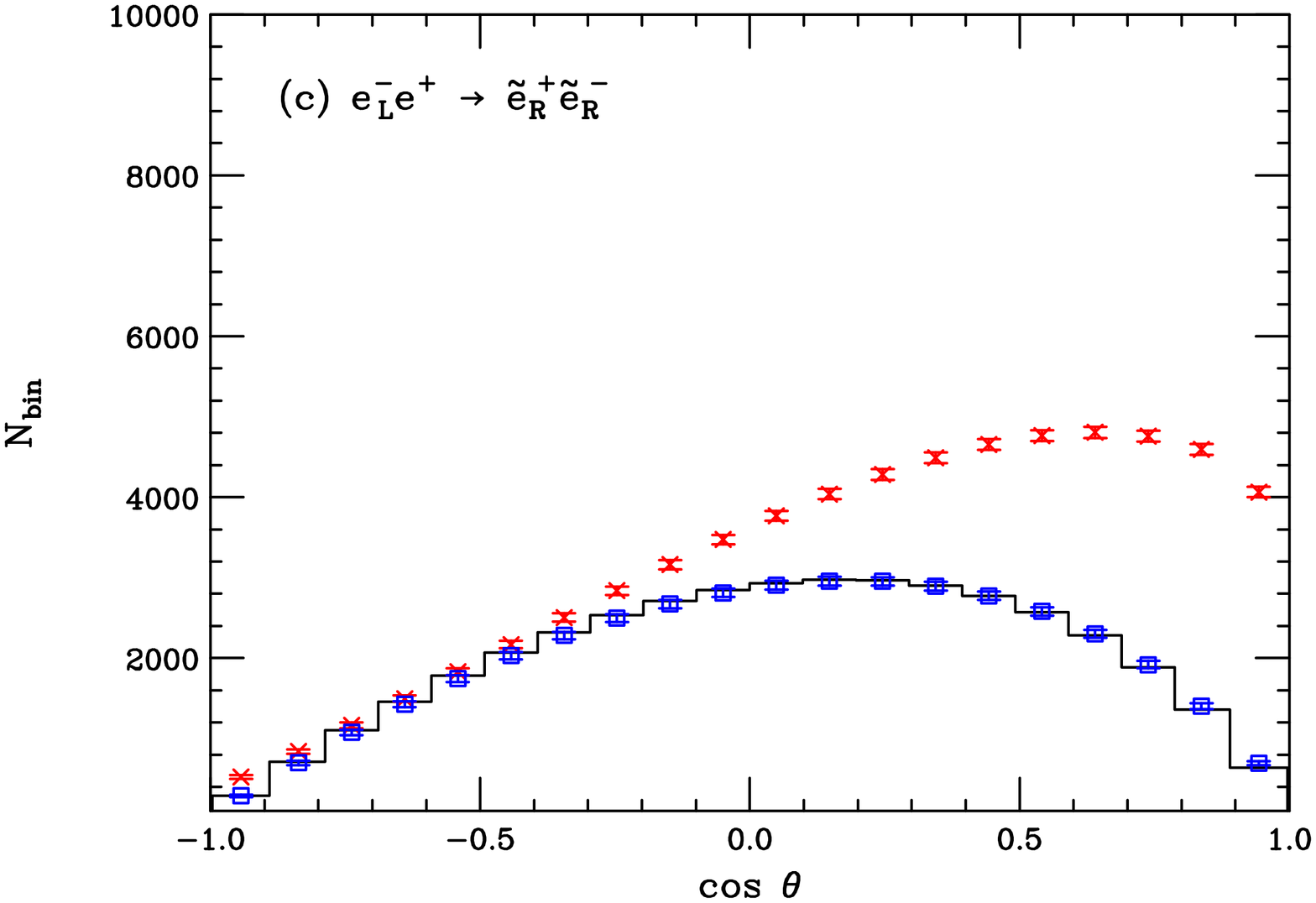,height=8cm,width=8cm,angle=90}
\hspace*{5mm}
\psfig{figure=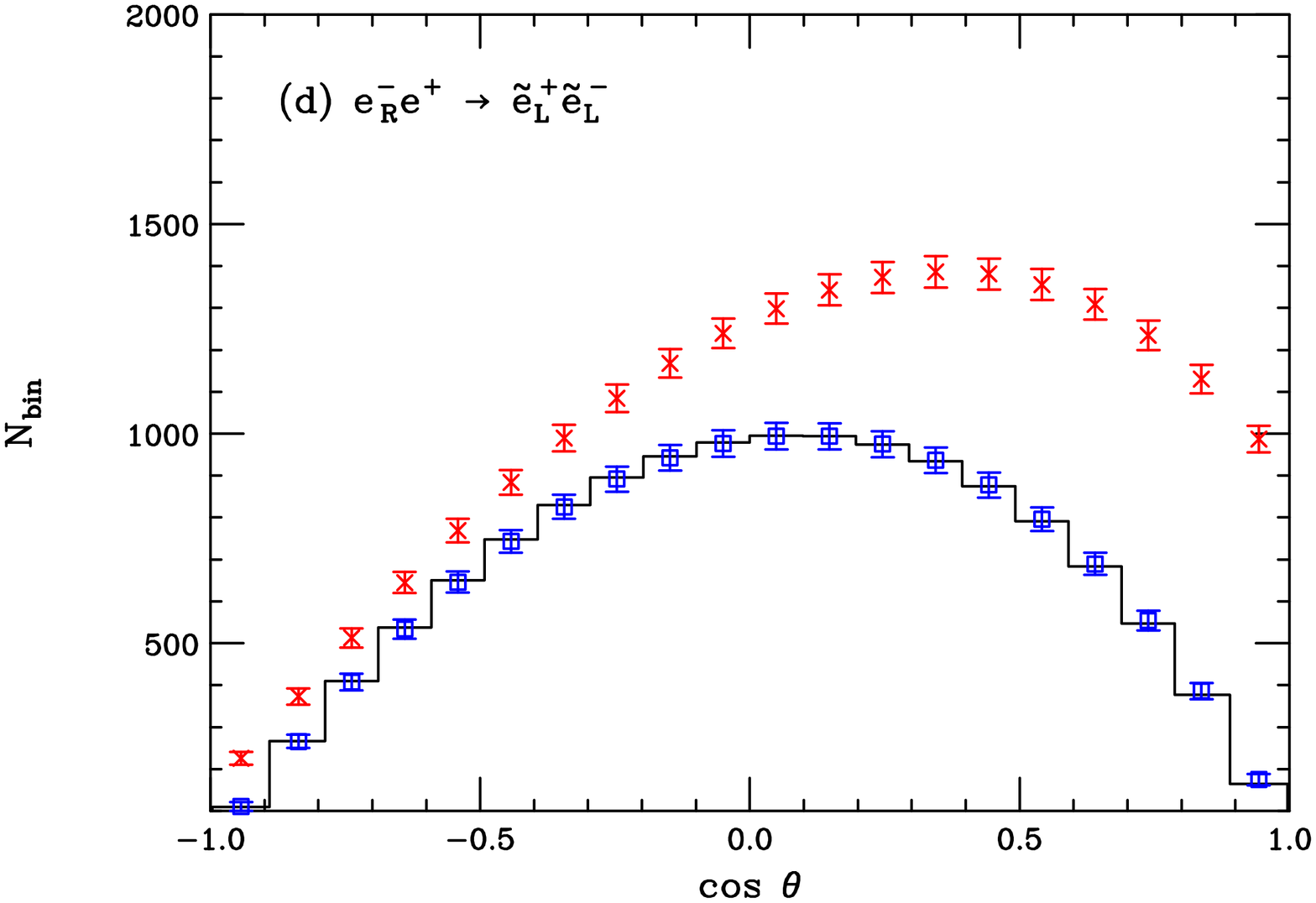,height=8cm,width=8cm,angle=90}}
\vspace*{0.25cm}
\centerline{
\psfig{figure=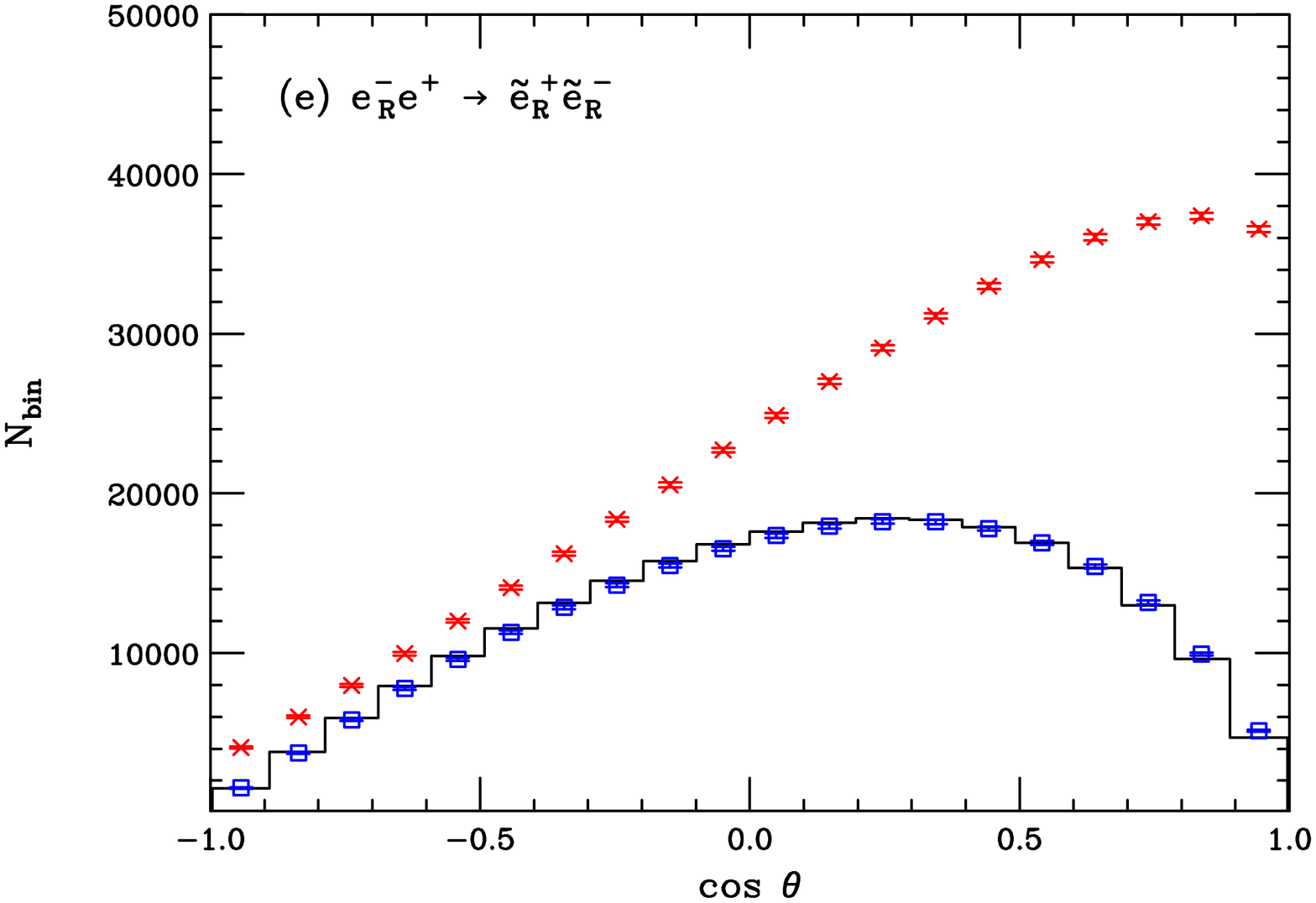,height=8cm,width=8cm,angle=90}}
\caption{(c-e) Polarized binned angular distributions for each helicity
configuration, taking an 80\% polarization of the initial electron
beam.  The solid histogram represents the $D=4$ bino-like model,
while the ``data'' points correspond to the effects of a supersymmetric
bulk with $\Lambda_c=1.5, 3.0, 6.0$ TeV from top to bottom.}
\label{8782_fig72}
\end{figure}

Next, we present in Fig. \ref{8782_fig71} and \ref{8782_fig72}
the number of events for the binned angular distribution
for each helicity configuration
with $80\%$ polarized electron beams for $\sqrt s = 500$ GeV
and $500\infb$ of integrated luminosity.  In each case, the solid
histogram corresponds to the $D=4$ bino-like supersymmetric model, while the
``data'' points represent the addition of the bulk graviton and gravitino KK
tower exchange for $\Lambda_c=1.5, 3.0, 6.0$ TeV from top to bottom.
As before, the contributions with $\Lambda_c=6.0$ TeV are only
distinguishable from the $D=4$ results in the case of $e^-_{L,R}e^+\to
\sl^\pm\sr^\mp$.  The error bars on the ``data'' points are statistical only.  
We see that in most of the reactions, the case of $\Lambda=3.0$ TeV 
leads to only a slight
increase in event rate in each bin, whereas for $e^-_{L,R}e^+\to\sl^\pm
\sr^\mp$, the t-channel  bulk gravitino KK exchange is significant, leading 
to observable deviations from the $D=4$ case even for $\Lambda_c=6.0$ TeV.

\nn

\nn

\begin{figure}[htbp]
\centerline{
\psfig{figure=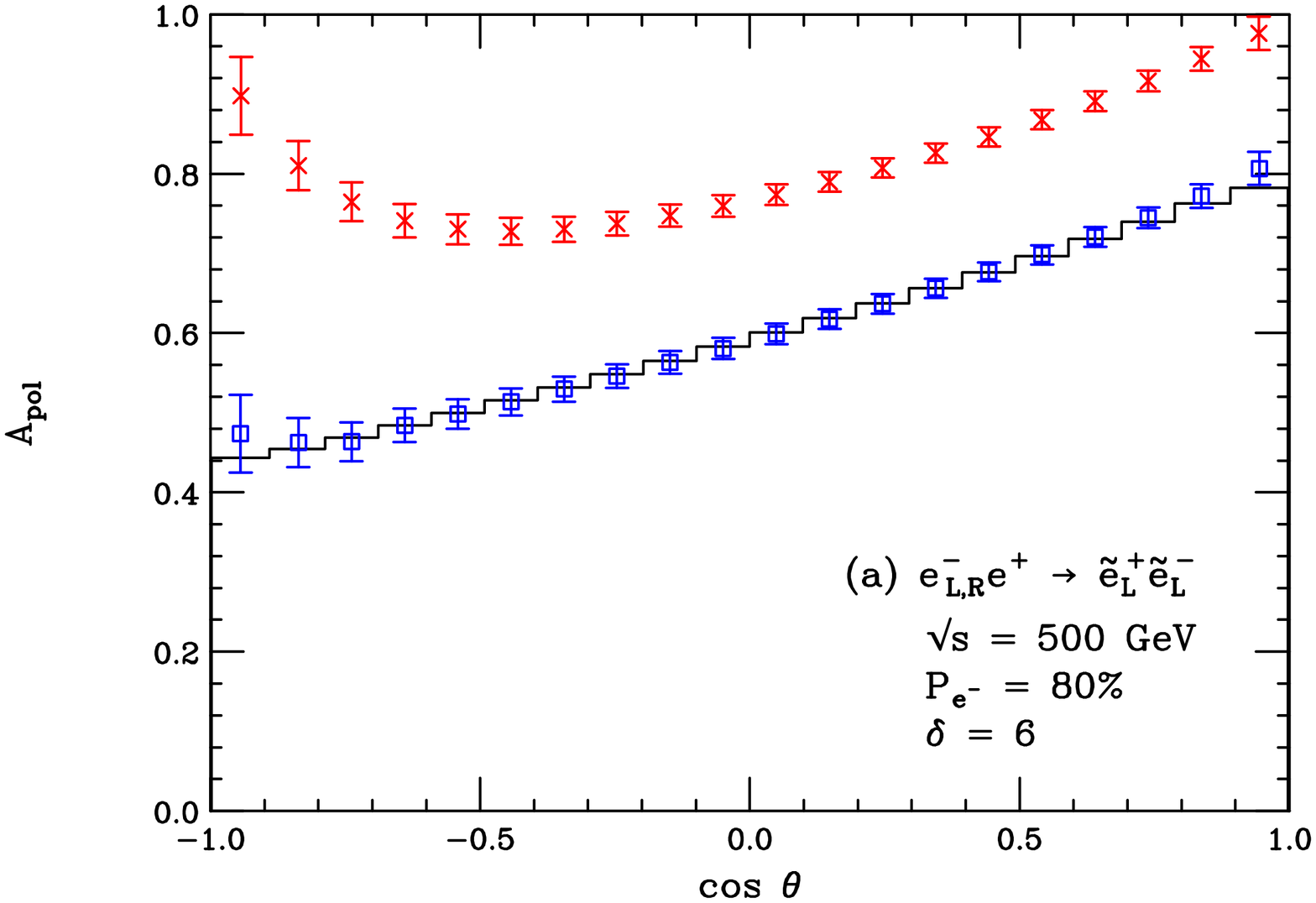,height=8.cm,width=8cm,angle=90}
\hspace*{5mm}
\psfig{figure=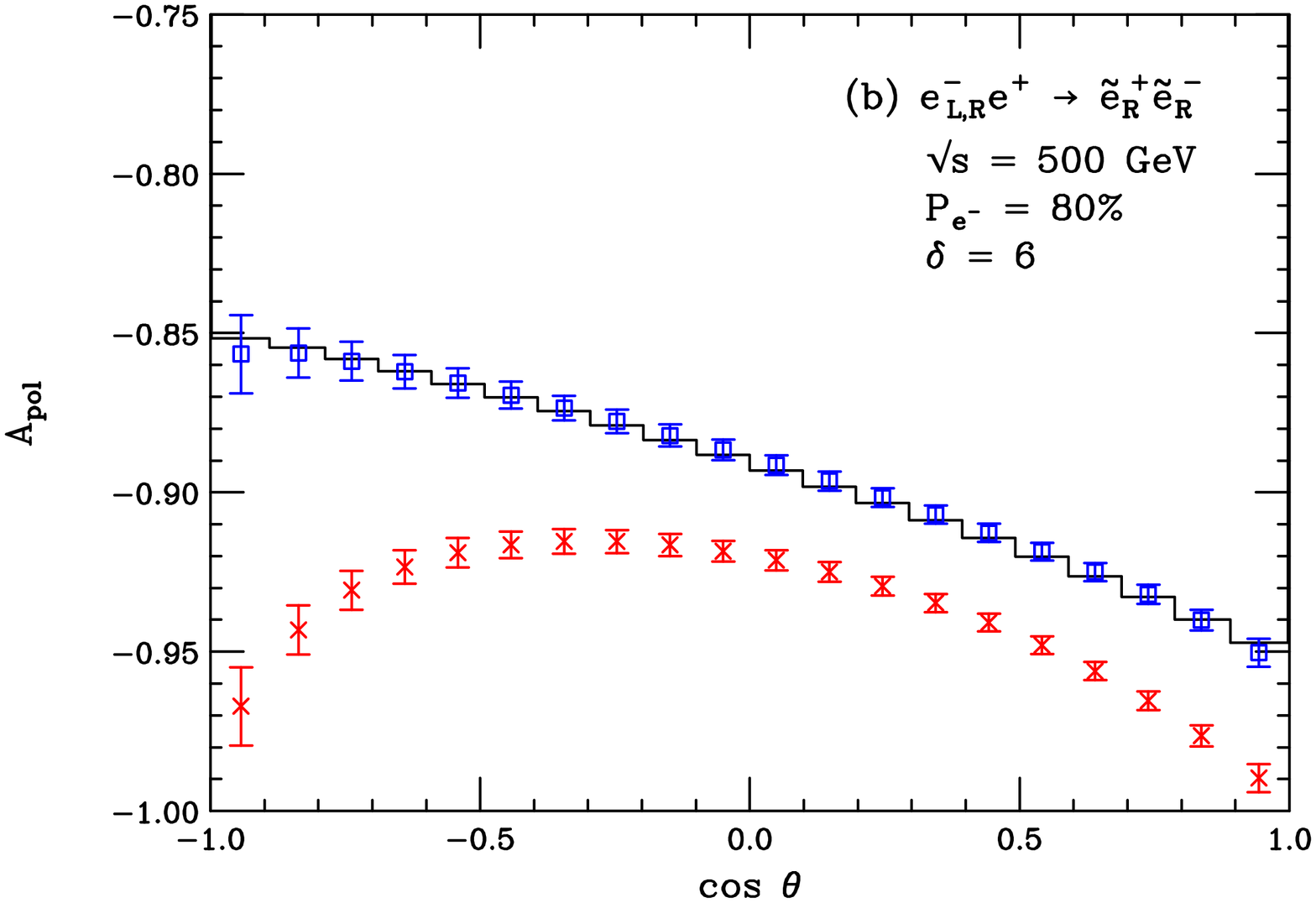,height=8.cm,width=8cm,angle=90}}
\caption{Polarization asymmetry defined in the text binned in 
$\cos\theta$.  The solid histogram represents the $D=4$ bino-like model,
while the two sets of ``data'' points include the contributions from
supersymmetric bulk with $\Lambda_c=1.5$ and 3.0 TeV.}
\label{8782_fig8}
\end{figure}

An interesting polarization asymmetry can be defined for the case of
$\sl^+\sl^-$ and $\sr^+\sr^-$ production.  It is given by
\begin{equation}
A_{\rm pol} = {d\sigma_L - d\sigma_R\over d\sigma_L + d\sigma_R}\,,
\end{equation}
where the left- and right-handed subscripts refer to the polarization of
the initial electron beam, \ie,
$d\sigma_i=d\sigma(e^-_ie^+\to\sl^-\sl^+\,, \sr^-\sr^+)/d\cos\theta$.  
This asymmetry is displayed in Fig. \ref{8782_fig8}, where the
solid histogram again represents our $D=4$ bino-like supersymmetric 
model and the ``data'' points are for a supersymmetric
bulk with $\Lambda_c=1.5$ and 3.0 TeV.  The error bars are again
statistical only and assume an integrated luminosity of 500 \infb.  The
electron beam polarization is taken to be 80\%.  We see that the
asymmetry varies substantially from its $D=4$ value with the addition 
of gravitino KK exchange, thus providing an additional signal for 
a supersymmetric bulk.

\nn

\begin{figure}[htbp]
\centerline{
\psfig{figure=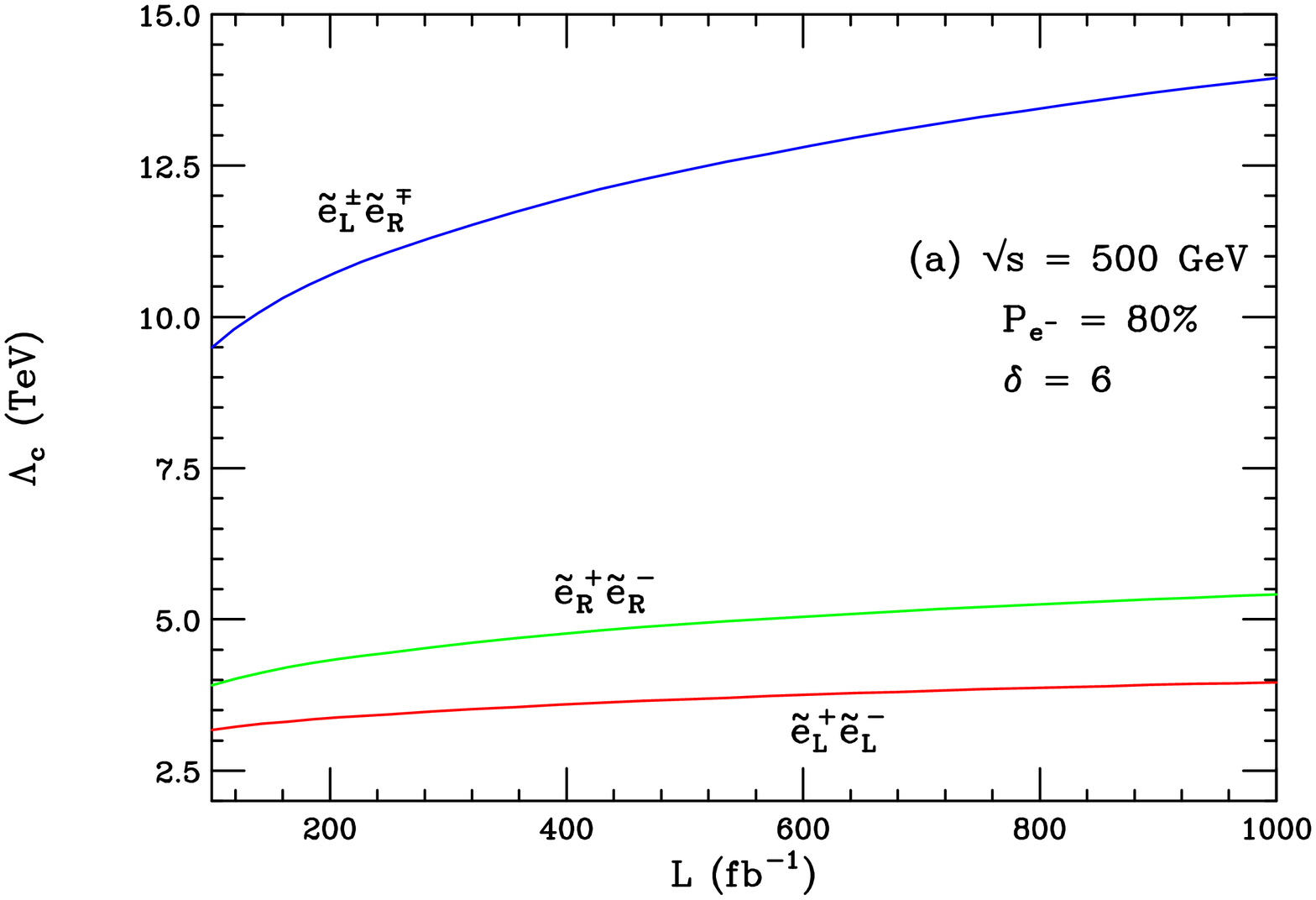,height=8.cm,width=8cm,angle=90}
\hspace*{5mm}
\psfig{figure=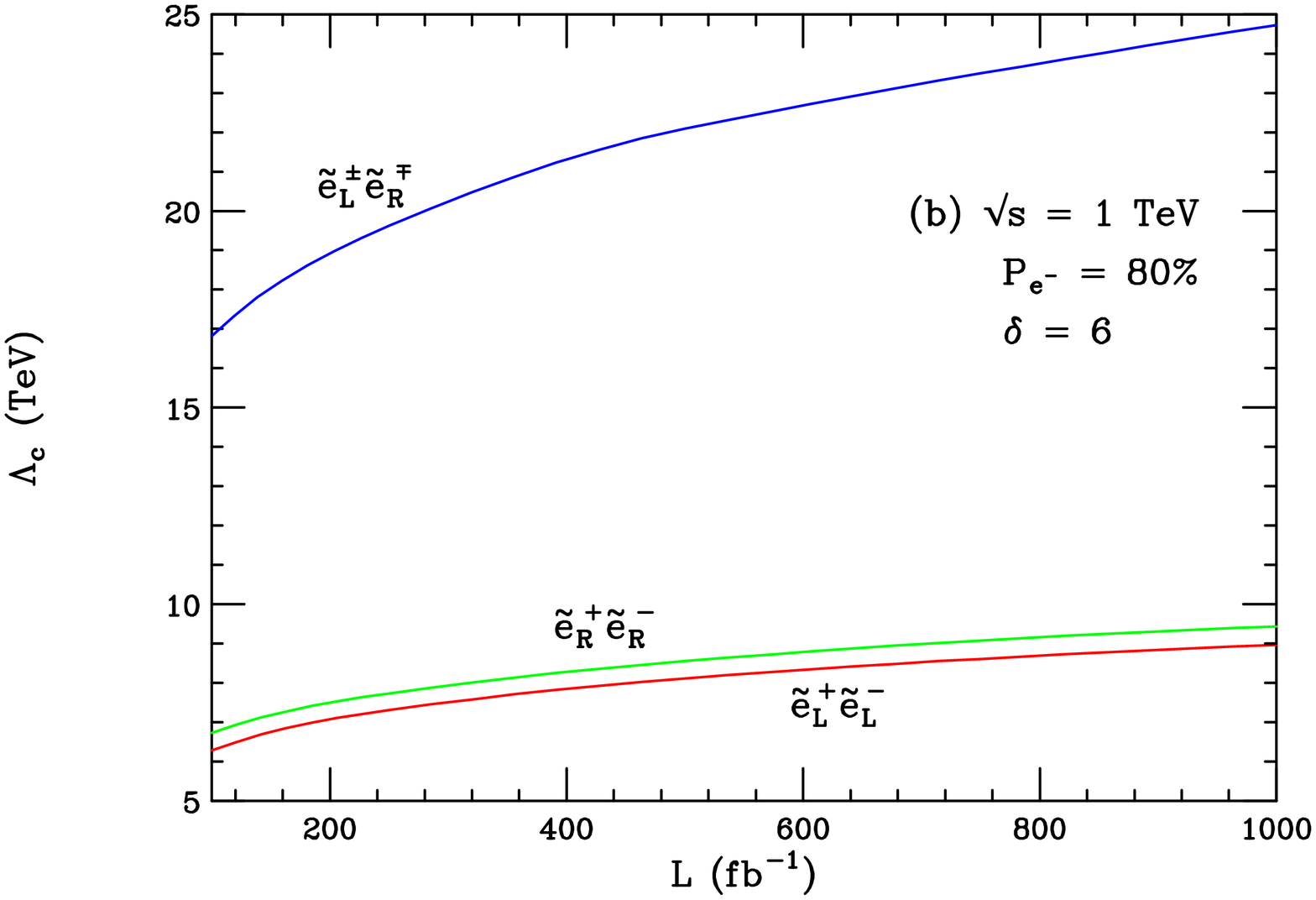,height=8.cm,width=8cm,angle=90}}
\caption{95\% C.L. search reach for $\Lambda_c$ in each production
channel as a function of
integrated luminosity for $\sqrt s= 0.5$ and 1.0 TeV.}
\label{8782_fig9}
\end{figure}

We now compute the potential sensitivity to the cut-off scale from 
selectron pair production using our sample case with a bino-like 
lightest neutralino state for purposes of demonstration.  We employ the 
usual $\chi^2$ procedure, taking
\begin{equation}
\chi^2 = \sum_{\rm bins} \left[ { \left({\frac{d\sigma}{d\cos\theta}}
\right)_{4d}
- \left({\frac{d\sigma}{d\cos\theta}}\right)_{10d} \over
\delta\left({\frac{d\sigma}{d\cos\theta}}\right) } \right]^2\,,
\end{equation}
where we include statistical errors only.  We sum over both initial left-
and right-handed electron polarization states, assuming $P_{e^-}=80\%$.
The resulting 95\% C.L. search for $\Lambda_c$ from each final
state, $\sl^+\sl^-\,, \sr^+\sr^-$, and $\sl^\pm\sr^\mp$, is given as a function
of integrated luminosity in Fig. \ref {8782_fig9} for $\sqrt s = 0.5$ and 1.0
TeV.  We see that for 500 \infb\ of integrated luminosity, corresponding
to design values, the search reach in the left- and right-handed selectron
pair production channels is given roughly by $\Lambda_c\simeq 6-10
\times\sqrt s$, which is essentially what is achievable for bulk graviton KK
exchange  in the reaction $e^+e^-\to f\bar f$ \cite{Hewett:1998sn}.
However, the $\sl^\pm\sr^\mp$ production channel yields an enormous
search capability with a 95\% C.L. sensitivity to $\Lambda_c$ of order
$25\times\sqrt s$ for design luminosity.  This process thus has the potential
to either discover a supersymmetric bulk, or eliminate the possibility of
supersymmetric large extra dimensions as being relevant to the hierarchy
problem.  We stress that there is nothing special about our choice of
supersymmetric parameters; our results will hold as long as selectrons 
are kinematically accessible to high energy \epem\ colliders.
We conclude that selectron pair production provides a very powerful
tool in searching for a supersymmetric bulk.

\nn

%

\section{Conclusions}

In summary, we have examined the phenomenological consequences of a 
supersymmetric bulk in the scenario of large extra dimensions.  We assumed
that supersymmetry is unbroken in the bulk, with gravitons and gravitinos
being free to propagate throughout the higher dimensional space, 
and that the SM and MSSM gauge and matter
fields are confined to a 3-brane.  Motivated by string theory, we
worked in the framework of $D=10$
supergravity, and found that the KK reduction of the bulk gravitinos
yields four Majorana spinors in four dimensions.  We then assumed that
the residual $N=4$ supersymmetry is broken near the fundamental scale 
$M_D$, with only $N=1$ supersymmetry surviving at the
electroweak scale.  

Starting with the $D=10$ action for this scenario, we expanded the
bulk gravitino into a KK tower of states, and determined the field
equation obeyed by the spin-3/2 KK excitations.  We then derived
the coupling of the bulk gravitino KK states to fermions and their
scalar partners on the brane.  We applied these results to a
phenomenological analysis by examining
the effects of virtual exchange of the  gravitino
KK tower in superparticle pair production.  We focused on the
reaction $\epem\to\tilde e^+\tilde e^-$ as this process is a
benchmark for collider supersymmetry studies.  Our numerical
analysis was performed in the framework of gauge mediated
supersymmetry breaking as it naturally affords a light
zero-mode gravitino.  However, our results do not depend on
the specifics of this particular model, with the exception of
the existence of a light zero-mode gravitino state.

Performing the sum over the KK propagators, we found that
the leading order contribution to this process arises from a
dimension-6 operator, and is independent of the zero-mode
mass.  This is in stark contrast to the virtual exchange of
spin-2 graviton KK states, which yields a dimension-8 operator
at leading order.  We thus found that the gravitino KK
contributions  substantially alter
the production rates and angular distributions for 
selectron pair production, and may essentially be isolated
in the $\sl^\pm\sr^\mp$ channel.  The resulting sensitivity
to the cut-off scale is tremendous, being of order
$20-25\times\sqrt s$.

We expect that the virtual exchange of gravitino KK
states in hadronic collisions will have somewhat less of an effect
in squark and gluino pair production than what we have found 
here.  The reason is that these
processes are initiated by both quark annihilation and gluon
fusion sub-processes, only one of which will be sensitive to tree-level
gravitino exchange for a given production channel.  The
sensitivity to the cut-off scale will then depend on the
relative weighting of the quark and gluon initial states.
In addition, t-channel gravitino
contributions will only be numerically relevant
for up- and down-squark production
due to flavor conservation;
hence their effect will be diluted by the production of the other
degenerate squark flavors and the relative weighting of the parton
densities.

Lastly, we note that virtual exchange of gravitino KK states
may also have a large effect on selectron pair production in
$e^-e^-$ collisions, which are tailor-made for  t-channel
Majorana exchanges.  High energy Linear Colliders thus provide
an excellent probe for the existence of  supersymmetric large
extra dimensions, and have the capability of discovering this
possibility or eliminating it as being relevant to the hierarchy
problem.

\section{Acknowledgments}

The authors would like to acknowledge useful discussions with
Hooman Davoudiasl, David London,
Maxim Perelstein, Michael Peskin, Frank Petriello, and Tom Rizzo.

\appendix
\section{Representation of the Dirac Algebra}
\label{appendix:rep}

A convenient representation of the Dirac algebra, which simplifies the
Kaluza-Klein decomposition, can be given in the $10$ dimensional
space-time as follows,
\begin{eqnarray} \label{10d:gamma:matrices}
  \Gamma^{\mu} \: & = & \: \gamma^{\mu} \: \otimes \: 1_{2} \otimes
  1_{4}\,,\nonumber \\
  \Gamma^{3+(2j-1)} \: & = & \: \gamma^{5} \: \otimes \: \sigma_{1}
  \: \otimes \: \alpha_{j}\,, \\
  \Gamma^{3+(2j)} \: & = & \: \gamma^{5} \: \otimes \: \sigma_{3}
  \: \otimes \: \beta_{j}\,,\nonumber
\end{eqnarray}
where the $\gamma^{\mu}$ are standard $4$ dimensional Dirac
matrices, $i,j = 1,2,3$ and $\mu=0,\ldots,3$.
Here $\alpha$ and $\beta$ are $4 \times 4$ matrices satisfying
\begin{subequations}
 \begin{align}
  \{ \alpha_{i} , \alpha_{j} \} \: &= \:
  \{ \beta_{i} , \beta_{j} \} \: = \:
  - 2 \delta^{ij}\,,\nonumber \\
  \{ \alpha_{i} , \beta_{j} \} \: &= \: 0\,,
 \end{align}
\end{subequations}
and the $\sigma$'s are standard pauli matrices satisfying
\begin{equation}
  \sigma^{i} \sigma^{j} \: = \: \delta^{ij} \: + \:
  i \epsilon^{ijk} \sigma^{k}\,,
\end{equation}
and $\gamma^{5}$ anticommutes with the $\gamma^{\mu}$
\begin{equation}
  \{ \gamma^{\mu} , \gamma^{5} \} \: = \: 0\,.
\end{equation}
The $\alpha$ and $\beta$ matrices can be represented as follows
\cite{Freedman:ra}
\begin{subequations}
  \begin{align}
    \alpha_{1} \: &= \:
    \left(
    \begin{matrix}
           0               &  \sigma_{1} \\
                -\sigma_{1}     &    0
    \end{matrix}
    \right)\,,
         &
    \alpha_{2} \: &= \:
    \left(
    \begin{matrix}
           0               &  -\sigma_{3} \\
                \sigma_{3}     &    0
    \end{matrix}
    \right)\,,
         &
         \alpha_{3} \: &= \:
    \left(
    \begin{matrix}
           i \sigma_{2}    &    0  \\
                0               &    i \sigma{2}
    \end{matrix}
    \right)\,,
  \end{align}\nonumber \\
  \begin{align}
    \beta_{1} \: &= \:
    \left(
    \begin{matrix}
           0               &  -i \sigma_{2} \\
                - i \sigma_{2}  &    0
    \end{matrix}
    \right)\,,
         &
    \beta_{2} \: &= \:
    \left(
    \begin{matrix}
           0               &  -1 \\
                1               &    0
    \end{matrix}
    \right)\,,
         &
         \beta_{3} \: &= \:
    \left(
    \begin{matrix}
           i \sigma_{2}    &    0  \\
                0               &    -i \sigma{2}
    \end{matrix}
    \right)\,.
  \end{align}
\end{subequations}
This representation of the Gamma matrices makes manifest the
decomposition to four dimensions, since $\Gamma^{\mu}$ for $\mu =
0,\ldots,3$ is the tensor product of four dimensional Gamma matrices
with an $8 \times 8$ identity acting on an internal index.
With these conventions, the space-like Gamma matrices are anti-hermitian,
while the time-like Gamma matrix is hermitian.
\section{Propagator for a Single Massive KK State}
\label{Propagator:sum}

To find the propagator for a single massive Kaluza-Klein state, we
invert the kinetic piece of the operator appearing in \eqref{action:standard:form}.
We solve
\begin{equation} \label{prop-eqn}
  \hat{O}^{\vec{n},\mu \nu} \: P^{\vec{n}}_{\nu \sigma} \: = \:
  i \: (k^2 - m_{\vec{n}}^2 ) \: \delta^\mu_\sigma
\end{equation}
for $P^{\vec{n}}_{\nu \sigma}$, where $\hat{O}^{\vec{n}, \mu \nu}$
is obtained by writing the Lagrangian for a single Kaluza-Klein state
in \eqref{action:standard:form} as
\begin{equation}
  \mathcal{L}^{\vec{n}} \: = \:
  \frac{1}{2} \bar{\Psi}_{\mu}
  \: \hat{O}^{\vec{n}, \mu \nu} \:
  \Psi_{\nu}\,.
\end{equation}
The propagator for the mode specified
by the vector of integers $\vec{n}$ is then given by
\begin{equation}
  \frac{P^{\vec{n}}_{\mu \nu}}{k^2 - m_{\vec{n}}^2 + i \epsilon}\,.
\end{equation}
We are free to drop the $i \epsilon$ convergence term as we are
interested in computing t-channel diagrams for which $t<0$, and so no
poles are encountered. To solve \eqref{prop-eqn}, we expand
$P^{\vec n}_{\mu \nu}$ as a linear combination of the standard set of
sixteen $4 \times 4$ matrices formed from antisymmetrized combinations
of the gamma matrices
$\{ 1, \gamma^{\mu}, \sigma^{\mu \nu}, \gamma^{\mu} \gamma^{5},
\gamma^{5}\}$, which form a basis for complex $4 \times 4$ matrices,
together with $\{ \eta_{\mu \nu}, k_{\mu}, k_{\nu} \}$, to generate the
correct tensor structure.  We then solve for the coefficients in this
expansion.  The result is
\begin{equation} \label{propagator}
  P^{\vec{n}, \mu \nu} \: = \:
  i \: \left( \slash{k} + m_{\vec{n}} \right)
  \left( \frac{k^{\mu} k^{\nu}}{m_{\vec{n}}^2} - \eta^{\mu \nu} \right)
  - \frac{i}{3}
  \left( \gamma^{\mu} + \frac{k^{\mu}}{m_{\vec{n}}} \right)
  \left( \slash{k} - m_{\vec{n}} \right)
  \left( \gamma^{\nu} + \frac{k^{\nu}}{m_{\vec{n}}} \right)
\end{equation}
and satisfies (on-shell) the standard condition $k_{\mu} P^{\mu \nu}=0$ together
with $\gamma_{\mu} P^{\mu \nu}=0$ which projects out the spin $\frac{1}{2}$
components.

\section{Summation of Kaluza-Klein States}

The summation over the Kaluza-Klein states contributing to the
propagator for the exchange of a
virtual gravitino KK tower is inherently more
complicated than in the case of  spin-$2$ exchange,
and leads to a quantitatively
different result. In addition, we must also
include the effects of the finite mass of
the zero-mode gravitino, $m_{0}$ \cite{Cheng:1999rx}.
We state here the basic result to leading order in the cut-off
$\Lambda_{c}$ for the case when $\Lambda_{c}^{2} \gg | t | ,
m_{0}^{2}$, and $\delta=6$ extra dimensions, which is the example considered
in the text.  The result is easily generalized to other values of
$\delta$.

The mass difference between neighboring KK states, 
$|\Delta m|\sim 1/R_c$, is much
smaller than the cut-off ($| \Delta m | \ll \Lambda_{c}$)
leading to a large number of states to be
summed over. The mass of the individual KK state is given by
\eqref{effective:Lagrangian:KK:diag}
\begin{equation} \label{KK:mass}
  m^{2}_{\vec{n}} \: = \:
  \frac{\vec{n} \cdot \vec{n}}{R_{c}^{2}}\,,
\end{equation}
where $R_{c}$ is the common compactification radius.
The near degeneracy of the KK state masses allows us to treat the
discrete population of  states on the lattice labeled by
the vector
$\vec{n}=(n_{1},\ldots,n_{\delta})$ in the continuum limit.
The number of KK states $dN$ in the thin shell between 
$m_{\vec{n}}^{2}$ and $m_{\vec{n}}^{2} + dm_{\vec{n}}^{2}$ is
\begin{equation}
  dN \: = \: \rho ( m_{\vec{n}}^{2} )
  d m_{\vec{n}}^{2}\,,
\end{equation}
with the density function
\begin{equation}
  \rho ( m_{\vec{n}}^{2} ) \: = \:
  \frac{(2 \pi R_{c})^{\delta} (m_{\vec{n}}^{2} - m_{0}^{2})^{\frac{\delta-2}{2}}}
       {(4 \pi)^{\frac{\delta}{2}} \Gamma ( {\frac{\delta}{2}} ) }\,.
\end{equation}

The coherent sum over these states is at the amplitude level
and involves terms of the form
\begin{equation} \label{integral:1}
  D_{\sigma} (t) \: \approx \: - i \:
  \int_{m_{0}^{2}}^{\Lambda_{c}^{2}} \: d m_{\vec{n}}^{2} \:
  \rho ( m_{\vec{n}}^{2} )
  \frac{F_{\sigma,\vec{n}}}{|t| + m_{\vec{n}}^{2}}\,,
\end{equation}
where the momentum exchange is in the t-channel with
$t < 0$.  Here, we have included an
explicit ultraviolet cut-off $\Lambda_{c}$ in order to regulate
this divergent integral.
As can be seen in Appendix \ref{Propagator:sum}, in the case of the
gravitino propagator there are four distinct classes of
$F_{\alpha, \vec{n}}$ to consider, given by
\begin{equation} \label{eqn:F}
  F_{\sigma,\vec{n}} \: = \: | m_{\vec{n}} |^{\sigma-2}\,,
\end{equation}
with $\sigma=0,1,2,3$.
Using \eqref{KK:mass}, \eqref{eqn:F} and the change of variables
$y=m_{\vec{n}} / \chi$, with $\chi \equiv \sqrt{|t|}$,
the integral \eqref{integral:1} can be put in the form
\begin{equation} \label{full:integral:6}
  D_{\sigma} (t) \: \approx \:
  \frac{-2 i (2 \pi R_{c})^{\delta} \chi^{\sigma + \delta -4}}
       {(4 \pi)^{\frac{\delta}{2}} \Gamma (\frac{\delta}{2})}
  \: \int_{m_{0}/\chi}^{\Lambda_{c}/\chi}
  \: dy \:
  \frac{(y^{2} - \frac{m_{0}^{2}}{\chi^{2}} )^{\frac{\delta - 2}{2}} y^{\sigma-1}}
       {1 + y^{2}}\,.
\end{equation}
This integral can be evaluated by making use of Appell's hypergeometric
function \\
$F_{1}(a;b_1,b_2;c;z_1,z_2)$
which generalizes hypergeometric series to two variables \cite{Bailey}.
It has the one dimensional integral representation
\begin{multline}
  \frac{\Gamma (a) \Gamma (c-a)}{\Gamma (c)} \:
  F_{1}(a;b_1,b_2;c;z_1,z_2) \: = \: \\
  \int_{0}^{1} \: u^{a-1} ( 1-u )^{c-a-1}
  (1-u z_{1})^{-b1} (1-u z_{2})^{-b2} \: du\,,
\end{multline}
and the series expansion
\begin{equation}
  F_{1}(a;b_1,b_2;c;z_1,z_2) \: = \:
  \sum_{m=0}^{\infty} \: \sum_{n=0}^{\infty} \:
  \frac{(a)_{m+n} (b1)_{m} (b2)_{n}}{m! n! (c)_{m+n} } \:
  z_{1}^{m} \: z_{2}^{n}\,,
\end{equation}
where $(a)_{n}$ is the Pochhammer symbol defined as
\begin{equation}
  (a)_{n} \: = \: a (a+1) \ldots (a+n-1)\,.
\end{equation}
For $\delta = 6$ extra dimensions, the result to leading
order in $\frac{\sqrt{|t|}}{\Lambda_c} \ll 1$ is
\begin{equation} \label{leading:order:6}
  D_{\sigma} \: \approx \:
  \frac{- i \: \pi^3 \: R_c^6 \: \Lambda_c^{\sigma + 2}}{\sigma + 2}\,.
\end{equation}
We note that at this order, the result is independent of the
mass of the zero-mode, so long as this mass is much smaller
than the cut-off scale.
The results for the various values of $\sigma$
are explicitly listed in Table 2. For $\delta=2$ extra dimensions,
the leading order result for $\sigma=3$ is
\begin{equation} \label{leading:order:2}
  D_{\sigma=3} \: \approx \:
  - 2 i \pi R_c^2 \: \Lambda_c \,.
\end{equation}
We note that the behavior of the sub-leading terms for $\delta=2$
is quite different than that displayed in Table 2 for the case of
$\delta=6$.
We do not rely on the above approximations, but evaluate
\eqref{full:integral:6} numerically in our analysis.
\begin{table} \label{integrals}
\centering
\begin{tabular}{|c|c|} \hline\hline
 $\sigma$ & $i \: D_{\sigma} \: ( \pi^{3} R_{c}^{6})^{-1}$
 \\ \hline
 0 & $\Lambda_c^{2} / 2$ \\ \hline
 1 & $\Lambda_c^{3} / 3$ \\ \hline
 2 & $\Lambda_c^{4} / 4$ \\ \hline
 3 & $\Lambda_c^{5} / 5$ \\ \hline \hline
\end{tabular}
\caption{Value of the integrals over the terms in the gravitino
propagator in six extra dimensions.}
\end{table}

An interesting feature of this result when applied to the gravitino propagator
is that it modifies its qualitative structure,
so that the summed propagator is dominated by a few
terms. This gives rise to the following ($\sigma=3$)
leading order behavior (for 6 extra dimensions)
\begin{equation}
  \sum_{\vec n} P^{\vec{n}, \mu \nu}
  \: \approx \:
  \frac{- i \pi^3 \Lambda_c^5 \: R_c^6}{5}
  \left (
    \eta^{\mu \nu} \: - \: \frac{1}{3} \gamma^{\mu} \gamma^{\nu}
  \right)\,.
\end{equation}
Following Han \etal, in \cite{Giudice:1998ck},
we take the relation between the compactification radius and the
cut-off scale to be
\begin{equation}
  R_{c}^{\delta} \: = \:
  \frac{8 \sqrt{\pi^{(2 - \delta)}} \: \Gamma (\frac{\delta}{2})}
       {\kappa^{2} \: \Lambda_{c}^{\delta + 2}}\,,
\end{equation}
with the gravitational coupling constant $\kappa = \sqrt{8 \pi G_N}$.

\def\IJMP #1 #2 #3 {Int. J. Mod. Phys. A {\bf#1},\ #2 (#3)}
\def\MPL #1 #2 #3 {Mod. Phys. Lett. A {\bf#1},\ #2 (#3)}
\def\NPB #1 #2 #3 {Nucl. Phys. {\bf#1},\ #2 (#3)}
\def\PLBold #1 #2 #3 {Phys. Lett. {\bf#1},\ #2 (#3)}
\def\PLB #1 #2 #3 {Phys. Lett. B {\bf#1},\ #2 (#3)}
\def\PR #1 #2 #3 {Phys. Rep. {\bf#1},\ #2 (#3)}
\def\PREV  #1 #2 #3 {Phys. Rev. {\bf#1},\ #2 (#3)}
\def\PRD #1 #2 #3 {Phys. Rev. D {\bf#1},\ #2 (#3)}
\def\PRL #1 #2 #3 {Phys. Rev. Lett. {\bf#1},\ #2 (#3)}
\def\PTT #1 #2 #3 {Prog. Theor. Phys. {\bf#1},\ #2 (#3)}
\def\RMP #1 #2 #3 {Rev. Mod. Phys. {\bf#1},\ #2 (#3)}
\def\ZPC #1 #2 #3 {Z. Phys. C {\bf#1},\ #2 (#3)}


\end{document}